\documentclass[twocolumn,floatfix,superscriptaddress]{revtex4-1}

\usepackage{amssymb}
\usepackage{amsmath}
\usepackage{bm}
\usepackage{amsfonts}
\usepackage{graphicx}
\usepackage{subfigure}
\usepackage{dcolumn}
\usepackage[colorlinks,citecolor=red,linkcolor=blue,urlcolor=blue,breaklinks=true]{hyperref}

\def\calP{{\cal P}}
\def\k{{\bf k}}
\def\S{{\bf S}}
\def\r{{\bf r}}
\def\sbar{\bar{s}}

\def\sdag{s^\dagger}
\def\tdag{t^\dagger}
\def\pbar{\bar{p}}

\def \d{\cdot}
\def\ua{\uparrow}
\def\da{\downarrow}

\allowdisplaybreaks

\begin{document}

\title[]{Triplon mean-field analysis of an antiferromagnet with degenerate Shastry-Sutherland ground states}

\author{Rakesh Kumar}
\email{rkumar@curaj.ac.in}
\affiliation{School of Physical Sciences, Jawaharlal Nehru University, New Delhi 110067, India}
\affiliation{Department of Physics, Central University of Rajasthan, Ajmer 305817, India}
\author{Brijesh Kumar}
\email{bkumar@mail.jnu.ac.in} 
\affiliation{School of Physical Sciences, Jawaharlal Nehru University, New Delhi 110067, India}

\begin{abstract}
  We look into the quantum phase diagram of a spin-$\frac{1}{2}$ antiferromagnet on the square lattice with degenerate Shastry-Sutherland ground states, for which only a schematic phase diagram is known so far. Many exotic phases were proposed in the schematic phase diagram by the use of exact diagonalization on very small system sizes. In our present work, an important extension of this antiferromagnet is introduced and investigated in the thermodynamic limit using triplon mean-field theory. Remarkably, this antiferromagnet shows a stable plaquette spin-gapped phase like the original Shastry-Sutherland antiferromagnet, although both of these antiferromagnets differ in the Hamiltonian construction and ground state degeneracy. We propose a sublattice columnar dimer phase which is stabilized by the second and third neighbor antiferromagnetic Heisenberg exchange interactions. There are also some commensurate and incommensurate magnetically ordered phases, and other spin-gapped phases which find their places in the quantum phase diagram. Mean-field results suggest that there is always a level-crossing phase transition between two spin gapped phases, whereas in other situations, either a level-crossing or a continuous phase transition happens.
\end{abstract}

\maketitle

\section{Introduction} % (fold)
\label{sec:introduction}

The emergence of exotic physical properties in strongly correlated systems is of great current interest~\cite{Lacroix2011,Balents2010,Chen2012,Zhang2009,Ji2007}. In particular, the spin-gapped systems with ground states made out of regularly arranged spin-singlets have been given a lot of attention in recent years~\cite{Misguich2013,Miyahara2003}. Such class of nonmagnetic ground states protects $SU(2)$ symmetry and shows exponentially decaying spin-spin correlations. The materials CuGeO$_3$ and SrCu$_2$(BO$_3$)$_2$ are two prototypical examples of gapped systems of the above type in 1D and 2D, respectively. Experimental investigations have shown that the first material is a good realization of the famous Majumdar-Ghosh (MG) model~\cite{Majumdar1969, Majumdar1969a,Hase1993, Castilla1995}, and the second material can be mapped on the notable Shastry-Sutherland (SS) model~\cite{Shastry1981,Kageyama1999}. Both of these models have two competing interactions that give rise to frustration. Moreover, the quantum mechanical nature of spins produces quantum fluctuations. These two effects play very important roles in the ground state selection, for example, dimer order of spin-singlets (regular pattern of spin-singlets on lattice bonds) emerges in both of these models. More precisely, the MG model consists of first and second neighbors antiferromagnetic Heisenberg exchange interactions on a spin-$\frac{1}{2}$ chain, and it shows an exact doubly degenerate dimerized ground state if the dimensionless parameter $\alpha(= J_1/J_2)$ is set to $2$~\cite{Majumdar1969, Majumdar1969a}. The SS model can be think of an extension of the MG model on square lattice with three-quarter of the diagonal bonds depleted out. The remaining diagonal bonds form an orthogonal motif on the square lattice. Ground state of the SS model is an exact state of orthogonal dimers at $\alpha=0.5$~\cite{Shastry1981}. It is shown that the orthogonal dimer state (or SS state) continues to exist for values of $\alpha\lesssim 0.7$~\cite{Miyahara1999}. A spin-gapped state consisting of the product of plaquettes has also been proposed earlier in the intermediate region $0.7\lesssim \alpha \lesssim 0.9$~\cite{Koga2000,Lauchli2002,Zhang2015}, and very recently it is confirmed by high-pressure inelastic neutron scattering experiments~\cite{Zayed2017}. These two important models have been generalized further~\cite{Kumar2002,Surendran2002,Schmidt2005,Danu2012} and inspired many other interesting exactly solvable constructions~\cite{Kumar2008,Kumar2009}.

In most generic models, an exact dimer ground state is elusive. The spin-$\frac{1}{2}$ $J_1$-$J_2$ Heisenberg antiferromagnet model on square lattice, one of the generic models, shows the existence of N\'eel and collinear (or stripe) magnetically ordered phases in it. Classically, the phase transition between these phases occurs at the `MG point' $\alpha=2$. But due to quantum fluctuations and frustration, a quantum paramagnetic phase appears in the intermediate region $0.4\lesssim J_2/J_1\lesssim 0.6$~\cite{Chakravarty1989, Sachdev1990, Poilblanc1991, Schulz1996, Oitmaa1996, Singh1999, Capriotti2001, Sushkov2001, Richter2010}. Although there are many different views on the nature of this phase, the majority of opinions suggest that it is a columnar-dimer (CD) singlet  phase~\cite{Sachdev1990, Poilblanc1991, Singh1999, Sushkov2001}. The type of quantum phase transition between a magnetically ordered phase (say, N\'eel) and a dimer phase (say, columnar) is also an important aspect of investigation. For instance, in the $J_1$-$J_2$ model, it is suggested by some researchers to be a weak first order transition~\cite{Kuklov2004, Sirker2006, Kruger2006}. However, within a field-theoretic framework, Senthil \emph{et al.}~\cite{Senthil2004, Senthil2004a} have proposed a generic scenario for a continuous quantum phase transition in such cases. They describe it not in terms of the usual order parameters as required in the Landau–Ginzburg theory but using the ``deconfined" quantum criticality notion in which the free spinons at the quantum critical point act as the essential degrees of freedom. With inspiring by this remarkable suggestion, many quantum spin models have been constructed and investigated to check this scenario~\cite{Batista2004, Sandvik2007, Gelle2008, Kumar2008, Kumar2009}. The signatures of such a deconfined quantum critical transition seem to have been observed in a spin-$\frac{1}{2}$ multiple-spin exchange model on the square lattice~\cite{Sandvik2007}. However, there is no finality yet on its occurrence in a wider set of model problems. There also exist counterexamples such as a ring-exchange model that breaks the Landau-Ginzburg framework, but shows a first order phase transition~\cite{Batista2004}.

Motivated by the models with exact dimer ground states and the hidden challenges in the recently proposed deconfined quantum criticality, G\'elle \emph{et al.} constructed a model which has an exact fourfold degenerate SS ground state (in contrast to the original SS model where there is no degeneracy)~\cite{Gelle2008}. This model consists of $J_1$-$J_2$ Heisenberg antiferromagnetic interactions and a multiple-spin exchange interaction $K$. It gives an \emph{exact} degenerate SS ground state when the Heisenberg interactions are turned off. Allowing nonzero values of $J_1$ and $J_2$, exactness nature of the SS states is destroyed but these states survive in the significant regions. Many other interesting magnetic and nonmagnetic phases emerge when energetically favorable conditions are met by tuning the interaction parameters. The authors studied this model numerically using finite-size exact diagonalization, and presented a schematic quantum phase diagram. A major problem of the exact diagonalization is that it cannot be done for systems in the thermodynamic limit. Moreover, the third neighbor Heisenberg exchange $J_3$ which naturally competes with the multiple-spin exchange interaction was missing in G\'elle \emph{et al.}'s investigations~\cite{Gelle2008}. Our present work focuses on these two issues. We systematic study the $J_1$-$J_2$-$J_3$-$K$ model using triplon mean-field theory, and present quantum phase diagrams in the thermodynamic limit.

% section introduction (end)

\section{Model} % (fold)
\label{sec:model}

\begin{figure}[htbp]
    \centering
        \includegraphics[width=0.49\textwidth]{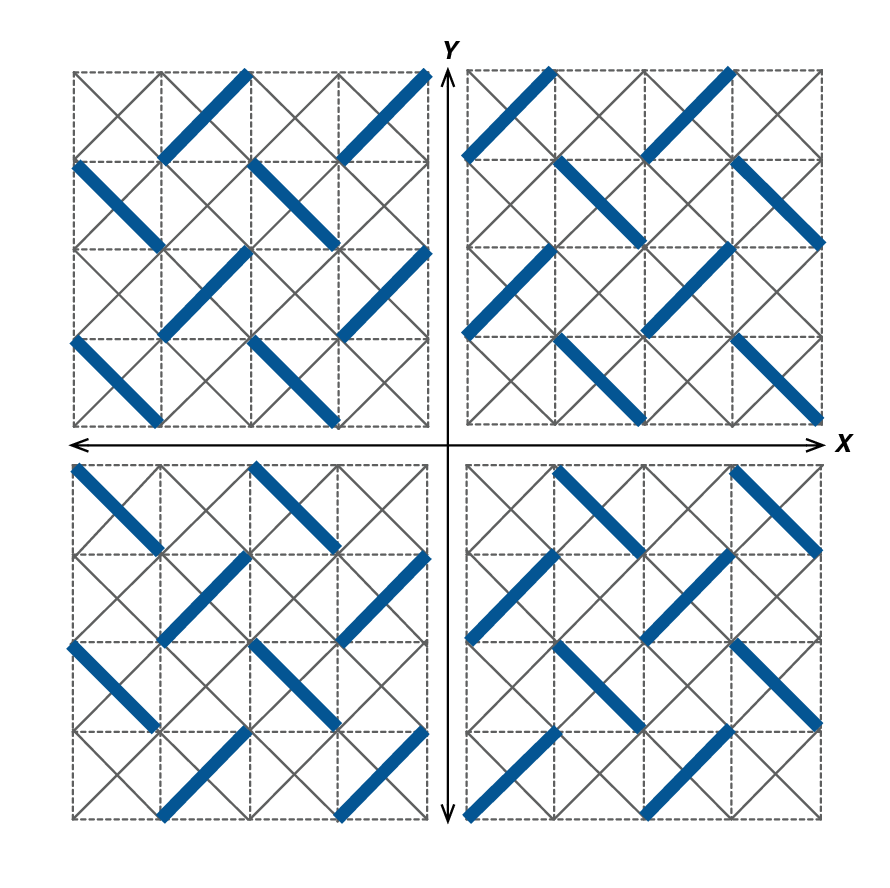}
    \caption{The four SS dimer states. These states spontaneously break the lattice translational symmetry and form an exact fourfold degenerate SS dimer ground state of the model~\eqref{eq:model_ss} when all $J$'s are kept to be zero.}
    \label{fig:ss4fold}
\end{figure}

We consider the following spin-$\frac{1}{2}$ Hamiltonian on square lattice
\begin{align}
  \label{eq:model_ss}
  \mathcal{H} = \mathcal{H}_{1}+\mathcal{H}_{2}+\mathcal{H}_{3}+\mathcal{H}_K
\end{align}
with
\begin{align}
  \mathcal{H}_{s} &= J_{s}\sum_{\langle i,j \rangle_s}\,\S_i\cdot\S_j,\quad s=1,2,3\\
  \mathcal{H}_K &= K \sum_{[i,j,k;l,m,n]}\, \calP_{3/2}(i,j,k)\,\calP_{3/2}(l,m,n),
\end{align}
where $J_1$, $J_2$, and $J_3$ are antiferromagnetic exchange couplings for first, second, and third nearest-neighbor bonds of the square lattice, respectively. The multiple-spin exchange interaction $\mathcal{H}_K$ is sum over all possible triangle-triangle interactions, and 
\begin{figure}[htbp]
\centering
\includegraphics[width=0.45\textwidth]{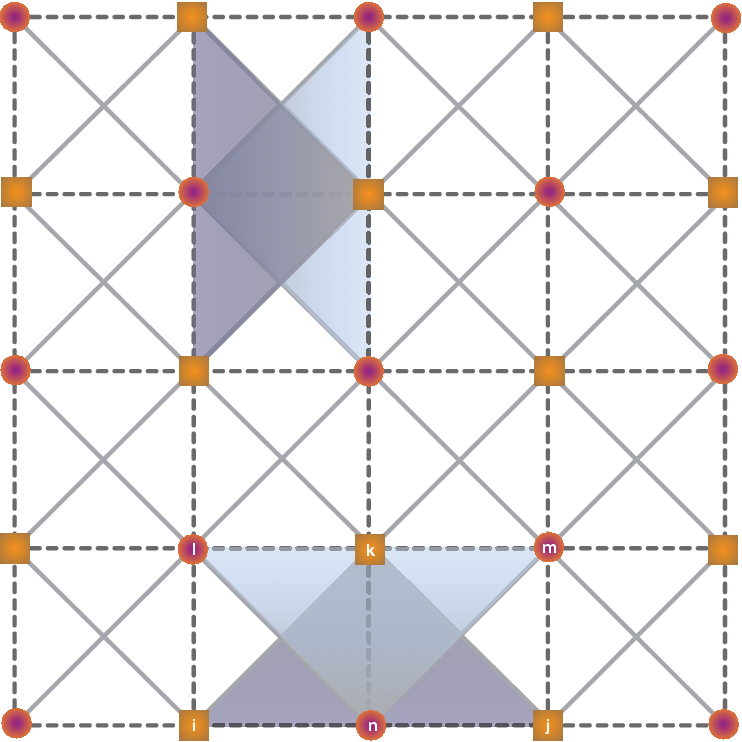}
\caption{Pictorial representation of the model Hamiltonian~\eqref{eq:model_ss}. The dashed lines between first and third neighboring sites represent $J_1$ and $J_3$ exchange couplings, respectively, and the solid lines denote $J_2$ exchange coupling. A set of two oppositely orientated triangles with vertices $i,j,k,l,m,n$ contributes $K \mathcal{P}_{3/2}(i,j,k)\mathcal{P}_{3/2}(l,m,n)$ in $\mathcal{H}_K$.} 
\label{fig:model_ss}
\end{figure}
each triangle-triangle interaction is a product of two projection operators $\calP_{3/2}(i,j,k)$ and $\calP_{3/2}(l,m,n)$ which are defined by the spins localized on the vertices $i,j,k,l,m,n$ of two oppositely oriented triangles, as shown in Fig.~\ref{fig:model_ss}. Opposite orientation of two triangles can be in the horizontal direction or in the vertical direction, and the $\mathcal{H}_K$ term includes all such orientations. The projection operator $\mathcal{P}_{3/2}$ is constructed in such a way that it projects a spin state with total spin three-half and annihilates otherwise, and it has the following form
\begin{equation}
  \label{eq:projector_operator}
    \mathcal{P}_{3/2}(i, j, k) = \frac{1}{3} \left( \S_{tot}^2 - \frac{3}{4} \right), \; \S_{tot} = \S_i + \S_j + \S_k.
\end{equation}
The exactly solvable limit of the model~\eqref{eq:model_ss} is achieved when all $J$'s are set to zero. In this limit, the eigenvalue equation for the Hamiltonian $\mathcal{H}$ gives four linearly independent degenerate ground states as shown by the four SS dimer states in Fig.~\ref{fig:ss4fold}.
By expanding $\mathcal{H}_K$ in Eq.~\eqref{eq:model_ss}, we obtain the following expression for the full Hamiltonian
\begin{align}
  \mathcal{H} &= \frac{1}{2}KL + J_1\sum_{\langle i,j \rangle}\S_i\cdot\S_j+ \sum_{s=2}^{3} \tilde{J}_s\sum_{\langle i,j\rangle_s}^{}  \S_i\cdot\S_j\nonumber \\
  &+\frac{4}{9}K \sum_{[i,j,k,l,m,n]} h(i,j,k) h(l,m,n)
  \label{eg:model_ss1},
\end{align}
where $L$ is the total number of lattice sites, $\tilde{J_2}=J_2+\frac{4}{3}\,K$, $\tilde{J_3}=J_3+\frac{2}{3}\,K$, and $h(i,j,k)=\S_i\cdot\S_j+\S_j\cdot\S_k+\S_k\cdot\S_i$. It is clear from the Eq.~\eqref{eg:model_ss1} that the $\mathcal{H}_K$ term in $\mathcal{H}$ renormalizes $J_2$ and $J_3$ coupling strengths but not $J_1$. For this reason we add $\mathcal{H}_3$ in the Hamiltonian given by G\'elle \emph{et al.}~\cite{Gelle2008}. 
% section model (end)

\section{Triplon mean-field theory} % (fold)
\label{sec:triplon_mean_field_theory}

A triplon mean-field theory is developed with respect to a quantum paramagnetic state in which spin-singlets are ``frozen" on some lattice units (for example, dimers or plaquettes)~\cite{Sachdev1990,Zhitomirsky1996,Kumar2008,Kumar2009,Kumar2010,Doretto2014}. Generally we consider homogeneity of spin-singlet objects in a quantum paramagnetic state, although, in principle, heterogeneous spin-singlet objects can also be taken. In this paper, only the first case is taken. First we choose a bilinear Hamiltonian and solve it exactly on a dimer or a plaquette lattice unit. We keep few lowest eigenstates of the Hamiltonian and discard the rest ones. Next it is hypothesized that some bosonic creation operators (singlet, triplet, etc.) produce these states out of vacuum. All these facilitate us to find a canonical mapping between spin operators and bosonic operators. An appropriate constraint in bosonic operators is also enforced to wipe out unphysical states. In the mean-field formulation, we assume Bose condensation of singlet operators on the frozen spin-singlets. Thus, triplet excitations (``triplons'') disperse in the singlet background and are responsible for a phase transition between a quantum paramagnetic phase and a magnetically ordered phase.

In the following two subsections, we formulate two kinds of triplon mean-field theories for Hamiltonian $\mathcal{H}$: dimer triplon mean-field theory~\cite{Sachdev1990,Kumar2008,Kumar2009,Kumar2010} and plaquette triplon mean-field theory~\cite{Zhitomirsky1996,Doretto2014}. These are devised by choosing a suitable dimerization and plaquettization on a lattice, respectively.

\subsection{Dimer triplon mean-field theory} % (fold)
\label{sub:dimer_triplon_mean_field_theory}

The $S=\frac{1}{2}$ Hamiltonian $\mathcal{H}_d=\S_1 \cdot \S_2 $ on a dimer yields four states: one singlet  state $|{ s} \rangle$ and three triplet states $|{ t}_x \rangle, |{ t}_y \rangle, |{ t}_z \rangle$. 
Suppose these four states are created by applying singlet and triplet creation operators on the vacuum $|0 \rangle$ as follows
\begin{subequations}
  \label{eq:bond_op}
  \begin{align}
  |{ s} \rangle &:=s^{\dag} |0 \rangle= \frac{1}{\sqrt{2}}\left(| \uparrow \downarrow\rangle  - |\downarrow \uparrow \rangle\right),  \\
  |{ t}_x \rangle &:={ t}_x^{\dag} |0 \rangle=  \frac{-1}{\sqrt{2}} \left(|\uparrow \uparrow\rangle - |\downarrow \downarrow\rangle\right), \\
   |{ t}_y\rangle &:={ t}_y^{\dag} |0 \rangle= \frac{i}{\sqrt{2}}\left(| \uparrow \uparrow\rangle +  |\downarrow \downarrow \rangle\right),   \\
  |{ t}_z\rangle &:={ t}_z^{\dag} |0 \rangle= \frac{1}{\sqrt{2}}\left(| \uparrow \downarrow\rangle + |\downarrow \uparrow \rangle\right). 
  \end{align}
\end{subequations}
The singlet and triplet operators, also known as \emph{bond-operators}~\cite{Sachdev1990}, satisfy the bosonic commutation relations
\begin{equation}
  [s, s^\dagger]=1,\; [t_ \alpha, t_ \beta^\dagger]=\delta_{\alpha \beta}, \; [s, t_ \alpha^\dagger]=0, \; \alpha,\beta=x,y,z.
\end{equation} 
In principle, the occupancy of bosons is not finitely restricted, therefore it causes inclusion of unphysical states. To avoid this, we impose the following constraint on each dimer
\begin{equation}
  \label{eq:dimer_constraint}
  s^\dag s + t_\alpha^\dag t_\alpha=1,
\end{equation}
where the repeated Greek index $\alpha$ is summed over. This constraint is another form of the \emph{completeness relation} on the basis states $|{ s} \rangle,|{ t}_x \rangle,|{ t}_y \rangle,|{ t}_z \rangle$, i.e.,  
\begin{equation}
  | s \rangle\langle s | + | t_{\alpha} \rangle\langle t_{\alpha} | = 1,
\end{equation}
where again the repeated index $\alpha$ is summed over. Clearly the last two equations assures us to have only physical states on a dimer unit. 

By calculating matrix elements of spin operators in the basis set $\left\{|{ s} \rangle,|{ t}_x \rangle,|{ t}_y \rangle,|{ t}_z \rangle \right\}$, we get the following mapping between spin operators and bond-operators
\begin{equation}
  \label{eq:spin_bo}
  S_{1,2}^{\alpha} = \frac{1}{2} \left(\pm s^\dag t_\alpha \pm t_\alpha^\dag s - 
  i\epsilon_{\alpha\beta\gamma} t_\beta^\dag t_\gamma \right),
\end{equation}
where $\epsilon_{\alpha\beta\gamma}$ is a totally antisymmetric tensor, and subscripts $1$ and $2$ refer spin labels on a dimer. Spin representation given in Eq.~\eqref{eq:spin_bo} is canonical, and thus the commutation relations of spin operators do not violate here. Now the bilinear spin-exchange on a dimer can be written as
\begin{equation}
  \label{eq:bilinear_spinexchange_dimer}
  \S_i \cdot \S_j = -\frac{3}{4}s^\dagger s + \frac{1}{4} t_ \alpha^\dagger t_ \alpha\approx -\frac{3}{4}\sbar^2 + \frac{1}{4} t_ \alpha^\dagger t_ \alpha,
\end{equation}
where the pre-factors $-\frac{3}{4}$ and $\frac{1}{4}$ are eigenvalues of the singlet state and the three triplet states of the $\mathcal{H}_d$, respectively. Similarly, the bilinear spin-exchange between two spins of different dimers can be approximately written as
\begin{align}
   \label{eq:bilinear_spinexchange_dimers}
\S_l(\r)\cdot\S_m(\r^\prime)&\approx (-)^{l+m} \frac{\sbar^2}{4} \left[ t_{\alpha}^\dagger(\r) t_{\alpha}(\r^\prime)+t_{\alpha}(\r)t_{\alpha}^\dagger(\r^\prime)\right.\nonumber\\
  &\left.+t_{\alpha}^\dagger(\r)t_{\alpha}^\dagger(\r^\prime)+t_{\alpha}(\r)t_{\alpha}(\r^\prime) \right],
\end{align}
where $l,m=1,2$ are spin labels, and $\r$ and $\r^\prime$ are the position vectors of dimers. In Eqs.~\eqref{eq:bilinear_spinexchange_dimer} and~\eqref{eq:bilinear_spinexchange_dimers}, we have replaced singlet operators by the condensate magnitude $\sbar$, that is, $\langle s^\dagger\rangle=\langle s \rangle=\sbar$. We ignore triplet-triplet interactions throughout in our analysis as they give a very little contribution~\cite{Sachdev1990}.

Below we work out dimer triplon mean-field theory for quantum paramagnetic states with one dimer and two dimers per unit cell. 

\subsubsection*{On a dimer state with one dimer per unit cell:} % (fold)
\label{ssub:on_a_dimer_state_with_one_dimer_unit_cell}

\begin{figure}[htbp]
    \centering
        \includegraphics[width=0.49\textwidth]{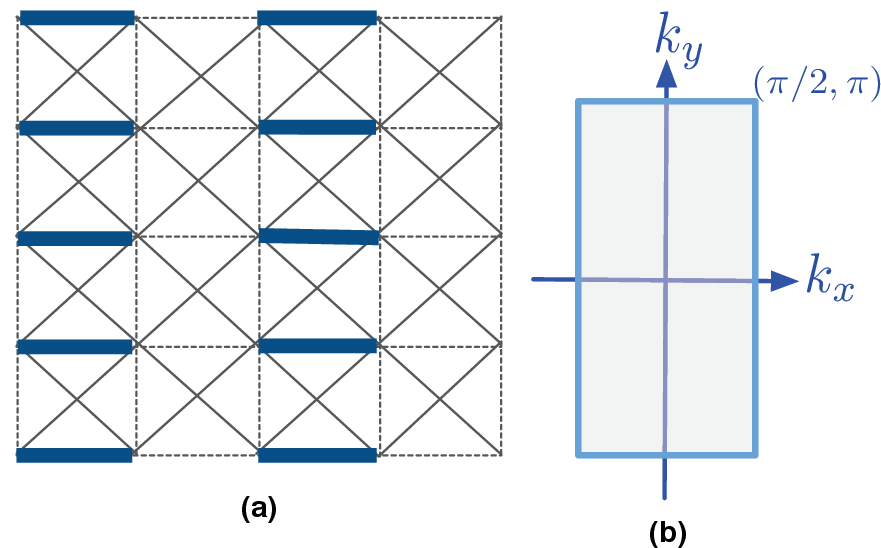}
    \caption{The columnar dimer state [in subfigure (a)] and its Brillouin zone [in subfigure (b)].}
    \label{fig:CDS}
\end{figure}

Many analytical and numerical works show that the spin-$\frac{1}{2}$ $J_1$-$J_2$ antiferromagnetic Heisenberg model has spin-gapped CD state (drawn in Fig.~\ref{fig:CDS}) in the intermediate region~\cite{Sachdev1990, Poilblanc1991, Singh1999, Sushkov2001}. 
Since the Hamiltonian~\eqref{eq:model_ss} contains $J_1$-$J_2$ exchange interactions, therefore we devise dimer triplon mean-field theory with respect to CD state. The objective is to see how this dimer state evolves or disappears when $J_3$  and $K$ are varied. The constraint~\eqref{eq:dimer_constraint} is applied globally by choosing the same chemical potential $\mu$ on each dimer: $-\mu\sum_\r\left[\sbar^2+\sum_\alpha\tdag_\alpha(\r)t_\alpha(\r)-1\right]$, where $\r$ denotes the position of a dimer. Now triplon mean-field Hamiltonian of the model~\eqref{eq:model_ss} with respect to CD state takes the following form
\begin{equation}
  \label{eq:mean_field_h_cds}
  \mathcal{H}_m=E_0+\mathcal{H}_k
\end{equation}
where
\begin{align}
E_0&=\frac{L}{2}\left[ K+\frac{1}{4}J_1-J_1\;\sbar^2-\frac{5}{2}\lambda+\lambda\;\sbar^2\right]\label{E:E0_CD},\\
H_k &=  \frac{1}{2}\sum_{\k,\,\alpha}\left[ \left(\lambda - \sbar^2\xi_\k\right)\left( t^\dag_{\k\alpha}t_{\k\alpha} + t_{-\k\alpha}t^\dag_{-\k\alpha}\right)  \right.\nonumber \\ & \left.- \sbar^2 \xi_\k \left( t^\dag_{\k\alpha}t^\dag_{-\k\alpha} +  t_{-\k\alpha}t_{\k\alpha}\right)\right].\label{E:Hmfk_CD}
\end{align}
In the above equations, $\lambda=\frac{1}{4}J_1-\mu$ is the effective chemical potential,  and $\xi_k=\frac{1}{2}J_1(\cos2k_x-2\cos k_y)+\tilde{J}_2(\cos2k_x\cos k_y+\cos k_y)-\tilde{J}_3(\cos2k_x+\cos2k_y)$.
After diagonalizing the mean-field Hamiltonian~\eqref{eq:mean_field_h_cds} using Bogoliubov transformation, we get
 \begin{equation}
\mathcal{H}_{m} = E_{0}+\sum_{\k,\,\alpha}\, E_{\k}\left(\gamma^{\dag}_{\k\alpha}\gamma^{ }_{\k\alpha} + \frac{1}{2} \right),
\label{E:Hdia_CD}
 \end{equation}
where $E_\k=\sqrt{\lambda(\lambda-2\bar{s}^2\xi_\k)}$ is known as the dispersion relation. Here, $\gamma$'s are Bogoliubov quasi-excitations. The saddle point equations of the ground state energy with respect to mean-field parameters $\sbar$, $\lambda$ lead to the self-consistent equations, and solution of which gives mean-field results. Detailed procedure of solving the self-consistent equations has been discussed in Ref.~\cite{Kumar2008}.

% subsubsection on_a_dimer_state_with_one_dimer_unit_cell (end)

\subsubsection*{On a dimer state with two dimers per unit cell:} % (fold)
\label{ssub:on_a_dimer_state_with_two_dimer_unit_cell}

\begin{figure}[htbp]
    \centering
        \includegraphics[width=0.49\textwidth]{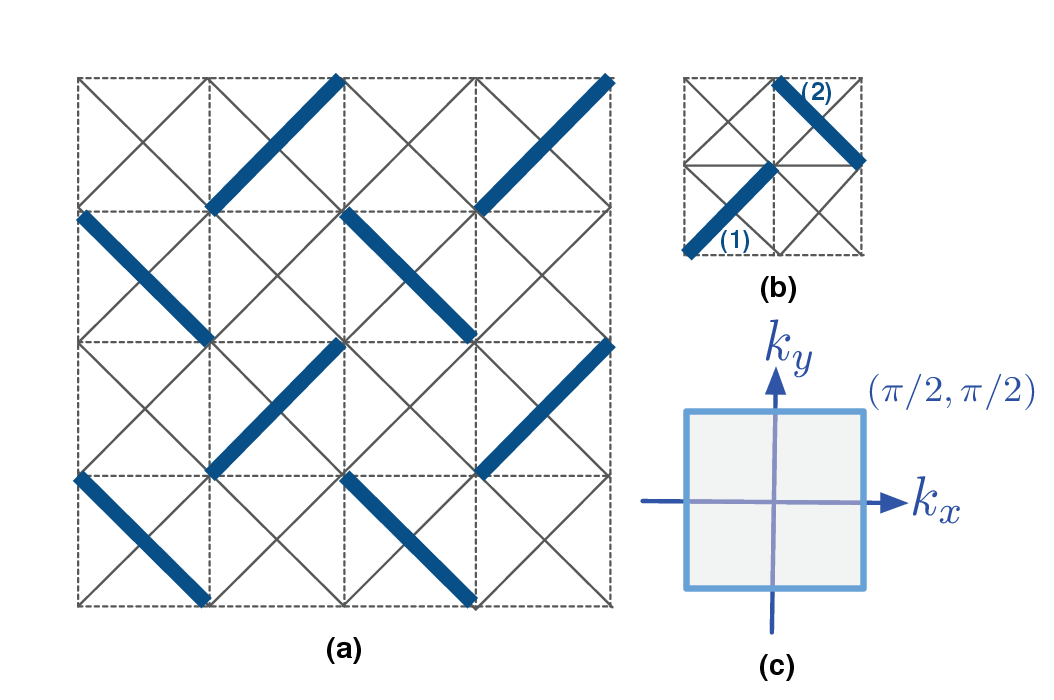}
    \caption{A SS dimer state is shown in Figure (a), and Figures (b) and (c) respectively display unit cell and Brillouin zone of the SS dimer state.}
    \label{fig:ss_dimer}
\end{figure}

\begin{figure}[h]
  \centering
    \includegraphics[width=.4\textwidth]{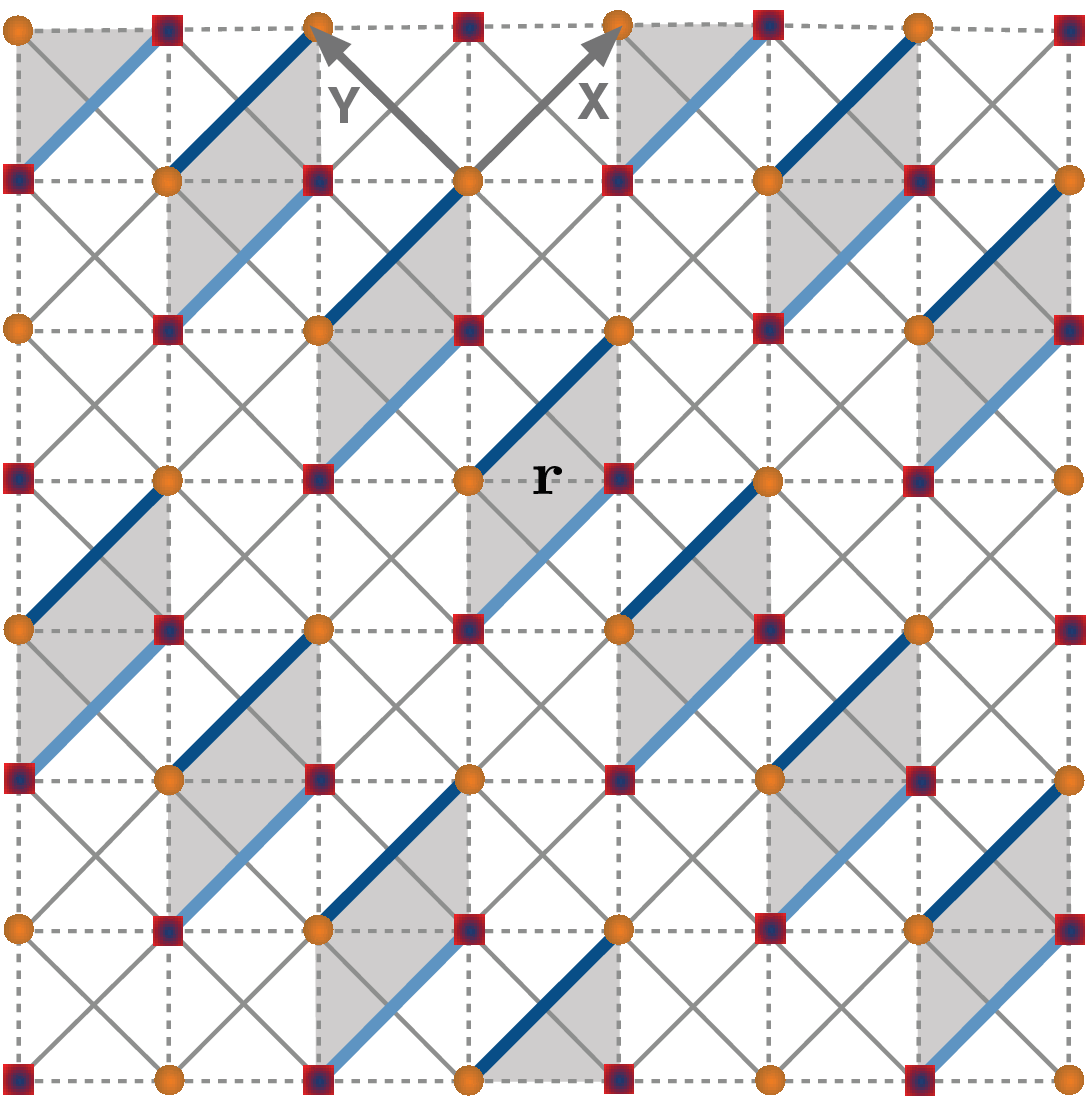}
  \caption{The sublattice columnar dimer state and its unit cell (a shaded region containing two dimers). The Brillouin zone of this state is same as for the CD state.}
  \label{fig:cds_sublattice}
\end{figure}

Since the model~\eqref{eq:model_ss} at exactly solvable limit has an exact fourfold degenerate SS ground state, it is natural to do triplon mean-field theory with respect to a SS dimer state (for example, as shown in Fig.~\ref{fig:ss_dimer}).
The unit cell of a SS dimer state contains two dimers [see Fig.~\ref{fig:ss_dimer}(b)]. Similarly, the sublattice columnar dimer (sCD) state, a CD state in which a dimer is formed between the same sublattice sites, shown in Fig.~\ref{fig:cds_sublattice} is also a possible choice of a quantum paramagnet state of the model~\eqref{eq:model_ss} in some parameter space. 
The reason is that $J_2$-$J_3$ model is topologically equivalent to the $J_1$-$J_2$ model. 

Since the unit cells of SS and sCD states have two dimers, therefore without loss of generality we choose two different sets of bond-operators, namely $\{s,t_{x},t_{y},t_{z}\}$ on one dimer and $\{p,q_{x},q_{y},q_{z}\}$ on the other dimer, where $s,p$ are singlet operators, and $t_{\alpha},q_{\alpha}$ are triplet operators. The singlet operators are replaced by their condensate amplitudes
\begin{equation}
  \langle s \rangle = \langle s^{\dagger} \rangle = \bar{s}, \quad \langle p \rangle = \langle p^{\dagger} \rangle = \bar{p}.
\end{equation}
Triplet operators together with singlet operators give two constraint equations
\begin{align}
  s^\dagger s+ t_{\alpha}^\dagger t_{\alpha}=1,\quad p^\dagger p+ q_{\alpha}^\dagger q_{\alpha}=1.
\end{align}
These constraints are imposed globally on all unit cells as follows
\begin{align}
  -\sum_{\r} \left\{\mu_s\left[\sbar^2+\sum_\alpha t_{\alpha}^{\dagger}(\r)t_{\alpha}(\r)-1\right]\right.\nonumber\\
  \left.+\mu_p\left[\pbar^2+\sum_\alpha q_{\alpha}^{\dagger}(\r)q_{\alpha}(\r)-1\right]\right\},
\end{align}
where $\mu_s$ and $\mu_p$ are the chemical potentials. Now the general form of the mean-field Hamiltonian with respect to both dimer states looks like
\begin{align}
  \label{eq:mean_field_hamiltonian_twodimers_unitcell}
  \mathcal{H}_{m}= E_0 + \frac{1}{2} \sum_{\k,\alpha}^{} \Psi^\dagger M \Psi,
\end{align}
where $\Psi^\dagger = \left( t_{\k \alpha}^\dagger\; q_{\k \alpha}^\dagger\;t_{-\k \alpha}\; q_{-\k \alpha}\right)$.  The constant $E_0$ and matrix $M$ with respect to SS dimer state take the following values 
\begin{align}
  E_0 &= \frac{L}{4} \left[2K + \frac{\tilde{J}_2}{2}-  \frac{5}{2} \left( \lambda_s + \lambda_p \right) \right.\nonumber \\
  &\left. -\tilde{J}_2 \left( \bar{s}^2 + \bar{p}^2 \right) + \lambda_s \bar{s}^2 + \lambda_p \bar{p}^2 \right],\label{eq:ss_e0}\\
  M &= \left( \begin{array}{cccc}
      \Lambda_s & 0 & \Delta_s & 0\\
      0 & \Lambda_p & 0 & \Delta_p \\
      \Delta_s & 0 & \Lambda_s & 0 \\
      0 & \Delta_p & 0 & \Lambda_p
    \end{array} \right),
\end{align}
where 
\begin{align}
  \Delta_s &=-\sbar^2 \zeta_0+\sbar^4\eta_0, \,\Delta_p =-\pbar^2\zeta_0 +\pbar^4\eta_0,\\
    \Lambda_s &= \Delta_s+\lambda_s, \,\Lambda_p = \Delta_p+\lambda_p,\\ \lambda_s&=\frac{1}{4}\tilde{J_2}-\mu_s,\, \lambda_p=\frac{1}{4}\tilde{J_2}-\mu_p,\\ \eta_0&=\frac{2}{3}K\cos2k_x\cos2k_y,\\ \zeta_0&=\frac{1}{2}\tilde{J_2}(\cos2k_x+\cos2k_y+\cos2k_x\cos2k_y)\nonumber\\
    &-\tilde{J_3}(\cos2k_x+\cos2k_y).
\end{align}
Similarly, for sCD state, the $E_0$ and $M$ are
  \begin{align}
    E_0 &= \frac{L}{4} \left[2K + \frac{\tilde{J}_2}{2}-  \frac{5}{2} \left( \lambda_s + \lambda_p \right)   \right.\nonumber \\
    &\left.-\tilde{J}_2 \left( \bar{s}^2 + \bar{p}^2 \right) + \lambda_s \bar{s}^2 + \lambda_p \bar{p}^2 +\epsilon_0\right],\\
    M &= \left( \begin{array}{cccc}
      \Lambda_s & \Gamma & \Delta_s & \Gamma\\
      \Gamma & \Lambda_p & \Gamma & \Delta_p \\
      \Delta_s & \Gamma & \Lambda_s & \Gamma \\
      \Gamma & \Delta_p & \Gamma & \Lambda_p
    \end{array} \right),
  \end{align} 
  where
  \begin{align}
    \Delta_s &= \sbar^2\zeta_0 + \eta_0, \,\Delta_p = \pbar^2\zeta_0 + \eta_0,\\
    \Lambda_s &=\Delta_s +\lambda_s,\,
    \Lambda_p = \Delta_p + \lambda_p,\\
    \Gamma &= \frac{1}{4}J_1 \bar{s}\bar{p} \left[ 1+\cos(k_y)\right.\nonumber\\
    &\left.-\cos(2k_x)-\cos(2k_x+k_y) \right],\label{eq:gamma_scd}\\
    \zeta_0 &=\frac{1}{2}\tilde{J}_2\left( 2\cos k_y - \cos 2k_x  \right) \nonumber \\
    &- \tilde{J}_3\left( \cos k_y  + \cos 2k_x \cos k_y\right), \\
    \eta_0 &=\frac{1}{3} K \sbar^2 \pbar^2 \left(-\cos k_y + \cos 2k_x \right.\nonumber\\ &\left.+ \cos 2k_x \cos k_y\right),\\
    \lambda_s&=\frac{1}{4}\tilde{J}_2  -\frac{1}{6}K\bar{p}^2   -\mu_s,\\ \lambda_p&=\frac{1}{4}\tilde{J}_2  -\frac{1}{6}K\bar{s}^2  -\mu_p,\\
    \epsilon_0&=\frac{1}{6}K \left( 5\sbar^2\pbar^2-\sbar^2 -\pbar^2\right).
  \end{align}
To diagonalize the mean-field Hamiltonian~\eqref{eq:mean_field_hamiltonian_twodimers_unitcell}, we need to solve the eigenvalue equation $\eta M \psi=\omega \psi$~\cite{Blaizot1986}. Here the matrix $\eta$ is given by
\begin{equation}
  \eta = \left( \begin{array}{cccc}
    1 & 0 & 0 & 0 \\
    0 & 1 & 0 & 0 \\
    0 & 0 & -1 & 0 \\
    0 & 0 & 0 & -1 
  \end{array} \right).
\end{equation} 
The eigenvalues of the matrix $\eta M$ are
\begin{align}
  \omega_1 &=-\omega_3=\sqrt{A_k+B_k},\\
  \omega_2 &=-\omega_4=\sqrt{A_k-B_k},
\end{align}
where
\begin{align}
  A_k &=\frac{1}{2} \left( \Lambda_s^2 - \Delta_s^2 \right) + \frac{1}{2} \left( \Lambda_p^2 - \Delta_p^2 \right),\\
  B_k&= \frac{1}{2}\sqrt{\left[ \left(\Lambda_s^2-\Delta_s^2\right) - \left( \Lambda_p^2-\Delta_p^2 \right) \right]^2+\epsilon},\\
  \epsilon &= 16 \Gamma^2 \left( \Lambda_s - \Delta_s \right)\left(\Lambda_p - \Delta_p\right).\label{eq:epsilon}
\end{align}
Now the mean-field Hamiltonian in diagonalized form can be written as
\begin{align}
  \mathcal{H}_{m}=E_0 +\sum_{\k \alpha}^{} &\left[ \omega_1 \left( \gamma_{\k \alpha}^\dagger \gamma_{\k \alpha} + \frac{1}{2}  \right) \right.\nonumber \\
  &\left.+ \omega_2 \left( \zeta_{\k \alpha}^\dagger \zeta_{\k \alpha} + \frac{1}{2}  \right) \right],
\end{align}
where $\gamma$ and $\zeta$ are quasi-particles. The saddle-point equations of ground state energy $E_g=E_0+ \frac{3}{2}\sum_{\k}^{} \left( \omega_1+\omega_2 \right)$  with respect to mean-field parameters ($\bar{s}^2$, $\bar{p}^2$, $\mu_s$, $\mu_p$) give self-consistent equations which are solved iteratively as explained in Ref.~\cite{Kumar2008}. It should be noted that $\omega_1=\omega_2$ for SS dimer state as in this case no mixing between $t_{\alpha}$ and $q_{\alpha}$ takes place.

% subsubsection on_a_dimer_state_with_two_dimer_unit_cell (end)

% subsection dimer_triplon_mean_field_theory (end)

\subsection{Plaquette triplon mean-field theory} % (fold)
\label{sub:plaquette_triplon_mean_field_theory}

\begin{figure}[htbp]
    \centering
        \includegraphics[width=0.35\textwidth]{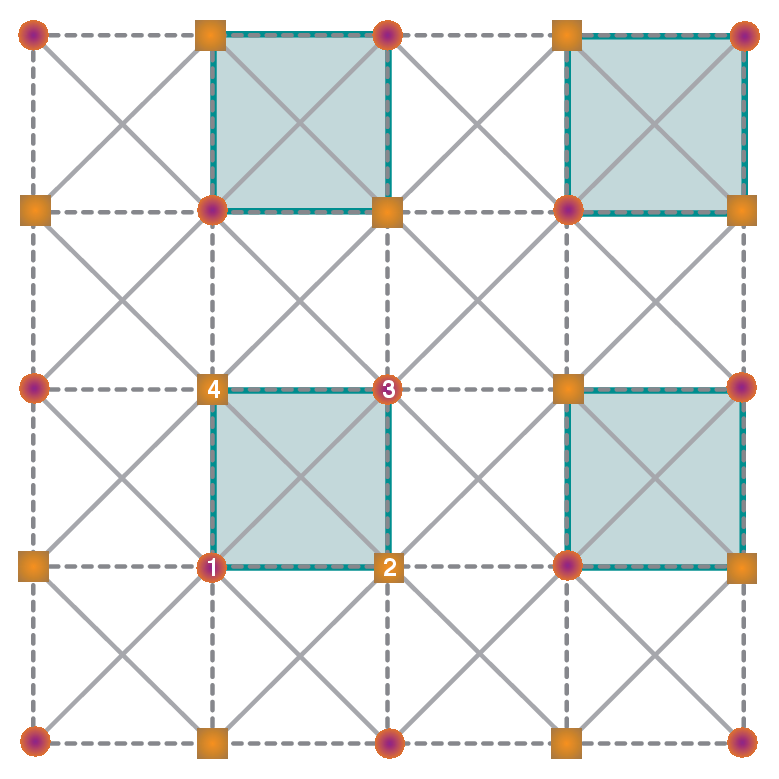}
    \caption{The plaquette resonating valence bond state. Its Brillouin zone is same as for the SS state.}
    \label{fig:plaquette}
\end{figure}

Here we work out triplon mean-field theory for Hamiltonian Eq.~\eqref{eq:model_ss} with respect to a nonmagnetic plaquette state shown in Fig.~\ref{fig:plaquette}. This state is chosen for two reasons: it is known to be exist as the ground state in a small region of $J_1$-$J_2$ model~\cite{Zhitomirsky1996,Capriotti2000,Mambrini2006} and also in the SS model as revealed by theoretical and experimental results~\cite{Koga2000,Lauchli2002,Zhang2015,Zayed2017}.  A plaquette state is a quantum paramagnetic state in which two spin-singlet dimers resonate on a four-spin block, and therefore it is also known as the \emph{plaquette resonating valence bond (pRVB) state}.

Consider a block of four spins in which spins are localized on sites $1,2,3,4$ as shown in Fig.~\ref{fig:plaquette}. The $J_1$-$J_2$ Heisenberg antiferromagnet on this small block takes the following form
\begin{align} \mathcal{H}_p&=J_1(\S_1\d\S_2+\S_2\d\S_3+\S_3\d\S_4+\S_4\d\S_1)\nonumber \\
  &+J_2(\S_1\d\S_3+\S_2\d\S_4).\label{eq:ss_hp}
\end{align}
Eigenspectrum of the Hamiltonian $\mathcal{H}_p$ can be found easily as follows. Define total spin operators $\S_{13}\equiv\S_1+\S_3$ and $\S_{24}\equiv\S_2+\S_4$ on the bonds $(13)$ and $(24)$, respectively. Then, total spin operator of the entire block of four spins can be written as $\S=\S_{13}+\S_{24}$. In terms of these operators, the $\mathcal{H}_p$ for spin-$\frac{1}{2}$ can be rewritten as
\begin{equation} \mathcal{H}_p=\frac{J_1}{2}(\S^2-\S_{13}^2-\S_{24}^2)+\frac{J_2}{2}(\S_{13}^2+\S_{24}^2-3),\label{eq:ss_hp_totalS}
\end{equation}
where the factor $3$ comes from $\S_1^2+\S_2^2+\S_3^2+\S_4^2$. Both $\S_{13}^2$ and $\S_{24}^2$ operators give eigenvalue $0$ or $1$ when applied on their eigenstates. Therefore, the four $S=\frac{1}{2}$ spins on the block $(1234)$ give two spin-singlets, three spin-triplets, and one spin-quintet. The complete eigenspectrum of spin-$\frac{1}{2}$ $\mathcal{H}_p$ is given in Table~\ref{tab:ss_hp_sprectrum}.

\begin{table}
    \caption{\label{tab:ss_hp_sprectrum}The eigenspectrum of the plaquette Hamiltonian $\mathcal{H}_p$. Here $[i,j]=\frac{1}{\sqrt{2}}\left(|\ua_i\da_j\rangle-|\da_i\ua_j\rangle\right)$ and $\{i,j\}=\frac{1}{\sqrt{2}}\left(|\ua_i\da_j\rangle+|\da_i\ua_j\rangle\right)$ denote singlet state and $S_z=0$ triplet state, respectively. The eigenstates corresponding to their eigenvalues are shown in the footnotes.}
    \begin{ruledtabular}
    \begin{tabular}{lccr}
        $S_{13}$ & $S_{24}$  & $S$ & Eigenenergy\\
        \hline
        0 & 0 & 0 & -$\frac{3}{2}J_2$\footnote{$[1,3]\otimes[2,4]=[1,2]\otimes[3,4]-[2,3]\otimes[4,1]$.}\\
        0 & 1 & 1 & -$\frac{1}{2}J_2$\footnote{
     $[1,3]\otimes|\da_2\da_4\rangle,\,[1,3]\otimes \left\{ 2,4 \right\},\,[1,3]\otimes|\ua_2\ua_4\rangle$.}\\
        1 & 0 & 1 & -$\frac{1}{2}J_2$\footnote{$|\da_1\da_3\rangle\otimes[2,4],\, \left\{ 1,3 \right\}\otimes[2,4],\,|\ua_1\ua_3\rangle\otimes[2,4]$.} \\
        1 & 1 & $\left\{\begin{array}{l}
            0 \\
            1 \\
            2
        \end{array}\right.$ & $\begin{array}{r}
             -2J_1+\frac{1}{2}J_2
    \footnote{$\frac{1}{\sqrt{3}}\left([1,2]\otimes[4,3]+[1,4]\otimes[2,3]\right)$.}\\   -J_1+\frac{1}{2}J_2\footnote{ $\frac{-1}{\sqrt{2}}|\da_1\da_3\rangle\otimes\{2,4\}+\frac{1}{\sqrt{2}}\{1,3\}\otimes|\da_2\da_4\rangle,\,
            \frac{1}{\sqrt{2}}|\ua_1\ua_3\rangle\otimes|\da_2\da_4\rangle -$\\\noindent $ \frac{1}{\sqrt{2}}|\da_1\da_3\rangle\otimes|\ua_2\ua_4\rangle,
    -\frac{1}{\sqrt{2}}|\ua_1\ua_3\rangle\otimes\{2,4\}+\frac{1}{\sqrt{2}}\{1,3\}\otimes|\ua_2\ua_4\rangle$.} \\
             J_1 +\frac{1}{2}J_2\footnote{ $|\da_1\da_3\rangle\otimes|\da_2\da_4\rangle, \, \frac{1}{\sqrt{2}}|\da_1\da_3\rangle\otimes\{2,4\}+\frac{1}{\sqrt{2}}\{1,3\}\otimes|\da_2\da_4\rangle,$\\\noindent$ 
            \sqrt{\frac{2}{3}}\{1,3\}\otimes \{2,4\}+\frac{1}{\sqrt{6}}|\da_1\da_3\rangle\otimes|\ua_2\ua_4\rangle + \frac{1}{\sqrt{6}}|\ua_1\ua_3\rangle\otimes|\da_2\da_4\rangle,$\\ \noindent$            \frac{1}{\sqrt{2}}|\ua_1\ua_3\rangle\otimes\{2,4\}+\frac{1}{\sqrt{2}}\{1,3\}\otimes|\ua_2\ua_4\rangle,\, |\ua_1\ua_3\rangle\otimes|\ua_2\ua_4\rangle$\,.}
        \end{array}$\\
    \end{tabular}                     
    \end{ruledtabular}
\end{table}

\begin{figure}[htbp]
  \centering
    \includegraphics[width=0.49\textwidth]{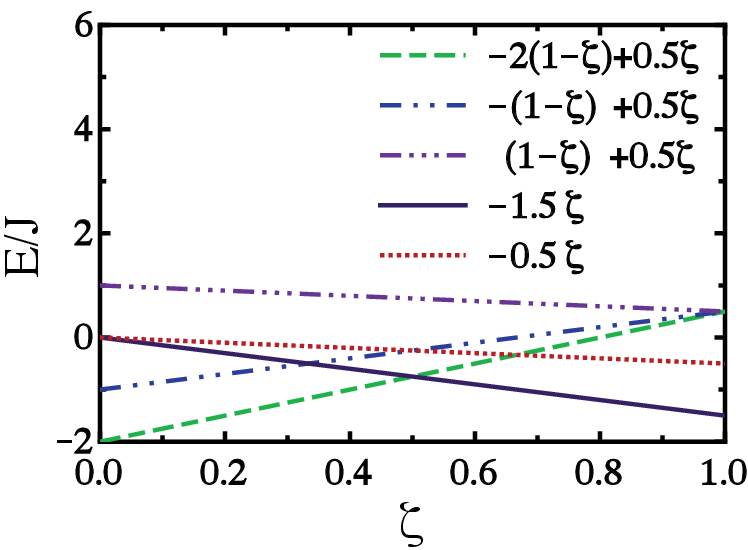}
  \caption{Eigenenergies of the Hamiltonian $\mathcal{H}_p$ in units of $J$. Here the varying parameter $\zeta$ comes from parametrization $J_1=(1-\zeta)J$, $J_2=\zeta J$.}
  \label{fig:eg_4spins}
\end{figure}

Next we do parametrization $J_1=(1-\zeta)J$, $J_2=\zeta J$ and plot eigenenergies of the Hamiltonian $\mathcal{H}_p$ in Fig.~\ref{fig:eg_4spins}.
If $\zeta < \frac{1}{3}$ (equivalently, $J_2/J_1<\frac{1}{2}$), the two lowest eigenvalues $-2J_1+\frac{1}{2}J_2$ and $-J_1+\frac{1}{2}J_2$ correspond to a spin singlet state and to three spin triplet states, respectively. We keep only these four states and develop the plaquette triplon mean-field theory. Imagine that the chosen states are emerging out of a vacuum by the application of boson creation operators
 \begin{equation}
   |s\rangle:=\sdag|0\rangle,\, |t_+\rangle:=\tdag_+|0\rangle,\,  |t_-\rangle:=\tdag_-|0\rangle,\,|t_z\rangle:=\tdag_z|0\rangle,
 \end{equation}
where
\begin{subequations}
      \begin{align}
          |s\rangle &= \frac{1}{\sqrt{3}}\left([1,2]\otimes[4,3]+[1,4]\otimes[2,3]\right),\qquad\\   |t_+\rangle&=\frac{-1}{\sqrt{2}}|\ua_1\ua_3\rangle\otimes\{2,4\}+\frac{1}{\sqrt{2}}\{1,3\}\otimes|\ua_2\ua_4\rangle,\\
  |t_-\rangle&=\frac{-1}{\sqrt{2}}|\da_1\da_3\rangle\otimes\{2,4\}+\frac{1}{\sqrt{2}}\{1,3\}\otimes|\da_2\da_4\rangle,\\
  |t_z\rangle&=\frac{1}{\sqrt{2}}|\ua_1\ua_3\rangle\otimes|\da_2\da_4\rangle - \frac{1}{\sqrt{2}}|\da_1\da_3\rangle\otimes|\ua_2\ua_4\rangle.
      \end{align}
\end{subequations}
Matrix elements of the spin operators, $\hat{S}_x$, $\hat{S}_y$, and $\hat{S}_z$, in the truncated subspace $\left\{|s\rangle, |t_+\rangle, |t_-\rangle , |t_z\rangle\right\}$ lead the following mapping
\begin{subequations}
  \label{eq:ss_spinop_plaquette}
  \begin{align}
      \hat{S}_m^z&=-\frac{(-)^m}{\sqrt{6}}\left(\tdag_z s+ \sdag t_z\right)
      + \frac{1}{4}\left(\tdag_+t_+-\tdag_-t_-\right),\\
      \hat{S}_m^+&=\frac{(-)^m}{\sqrt{3}}\left(\tdag_+ s+ \sdag t_-\right)
      + \frac{1}{2\sqrt{2}}\left(\tdag_+t_z-\tdag_zt_-\right),\qquad\\
      \hat{S}_m^-&=\frac{(-)^m}{\sqrt{3}}\left(\tdag_- s+ \sdag t_+\right)
      + \frac{1}{2\sqrt{2}}\left(\tdag_zt_+-\tdag_-t_z\right),
  \end{align}
\end{subequations}
where $m(=1,2,3,4)$ denotes vertex label in a plaquette. To fix the number of bosons on each plaquette, the constraint $\sdag s+ \tdag_+t_++\tdag_zt_z+\tdag_-t_-=1$ is imposed therein. Defining the operators $t_x$ and $t_y$ as follows
\begin{align}
    t_x := \frac{-1}{\sqrt{2}}\left(t_++t_-\right),\quad
    t_y := \frac{1}{\sqrt{2}i}\left(t_+-t_-\right).
\end{align}
Now the constraint equation becomes $\sdag s+ \tdag_xt_x+\tdag_yt_y+\tdag_zt_z=1$, and the spin-operators shown in Eq.~(\ref{eq:ss_spinop_plaquette}) can be compactly written as
\begin{equation}
    \hat{S}_m^\alpha=-\frac{(-)^m}{\sqrt{6}}\left(\tdag_\alpha s+ \sdag t_\alpha\right)-\frac{i}{4}\epsilon_{\alpha\beta\gamma}\,\tdag_\beta t_\gamma,
\end{equation}
where $\alpha, \beta,\gamma=x,y,z$, and $\epsilon_{\alpha\beta\gamma}$ is the Levi-Civita antisymmetric tensor.

In the quadratic triplon mean-field theory, we ignore mixing of different triplet modes and higher order triplet interactions as they give a very little contribution~\cite{Sachdev1990}. Under these assumptions, the $J_1$-$J_2$ Heisenberg antiferromagnet model on a plaquette positioned at $\r$ is written as
\begin{align}
  \mathcal{H}_p\left(\r\right)&=\left(-2J_1+\frac{1}{2}J_2\right)\sdag(\r)s(\r)\nonumber \\ &+\left(-J_1+\frac{1}{2}J_2\right)\sum_\alpha\tdag_\alpha(\r) t_\alpha(\r),
\end{align}
and the exchange-interaction between spins of two different plaquettes can be approximated as
\begin{align}
&\S_m(\r)\cdot\S_n(\r^\prime)\approx\frac{(-)^{m+n}}{6}\sbar^2 \sum_\alpha\left[t_\alpha(\r)\tdag_\alpha(\r^\prime) \right.\nonumber \\
&\left. +\tdag_\alpha(\r)t_\alpha(\r^\prime) + t_\alpha(\r)t_\alpha(\r^\prime)+\tdag_\alpha(\r)\tdag_\alpha(\r^\prime)\right],\qquad
\end{align}
where $\r$ and $\r^\prime$ are the position vectors of two different plaquettes, and $m,n=1,2,3,4$ are the vertex labels.

Let us assumed that the singlet bosons condense on the plaquettes. Under this assumption, we can replace singlet operators by their condensation amplitude $\sbar$, i.e., $\langle\sdag\rangle=\langle s\rangle=\sbar$. Next the constraint is imposed globally, that is,  $-\mu\sum_\r\left[\sbar^2+\sum_\alpha\tdag_\alpha(\r)t_\alpha(\r)-1\right]$ on each plaquette. The mean-field Hamiltonian can now be written as
\begin{align}\label{eq:ss_mfh_pl}
\mathcal{H}_{m} &= E_0+\sum_{\k,\,\alpha}\left[\Lambda_\k\left(t_{\k\alpha}\tdag_{\k\alpha}+\tdag_{\k\alpha}t_{\k\alpha}\right)\right.\nonumber\\
&\left.+\Delta_\k\left(t_{\k\alpha}t_{-\k\alpha}+\tdag_{\k\alpha}\tdag_{-\k\alpha}\right)\right],    
\end{align}
where
\begin{align} \frac{E_0}{L}&=-\frac{1}{4}J_1+\frac{1}{8}\tilde{J}_2+\frac{1}{2}K-\frac{5}{4}\lambda+\frac{1}{2}\lambda\,\sbar^2\nonumber\\&-\sbar^2\left(\frac{1}{4}J_1+\frac{1}{12}K-\frac{1}{36}K\,\sbar^2\right),\\
    \Lambda_\k&=\lambda+\frac{1}{9}K\,\sbar^2+\Delta_\k,\\   \Delta_\k&=\frac{1}{3}\sbar^2\left[\left(-J_1+\tilde{J}_2+2\tilde{J}_3+\frac{4}{9}K\,\sbar^2\right)\right.\nonumber \\ &\left.\left(\cos 2k_x+\cos 2k_y\right)+\tilde{J}_2\cos 2k_x\,\cos 2k_y\right].\quad
\end{align}
In the above equations, the definition of renormalized effective potential, $\lambda=\frac{1}{2}\left(-J_1+\frac{1}{2}\tilde{J_2}-\mu\right)$, is used. Finally, we diagonalize the mean-field Hamiltonian Eq.~(\ref{eq:ss_mfh_pl}) by Bogoliubov transformation. The diagonalized mean-field Hamiltonian is then reduced to
\begin{equation}  \mathcal{H}_{m}=E_0+2\sum_{\k,\,\alpha}E_\k\left(\gamma^\dagger_{\k\alpha}\gamma_{\k,\,\alpha}+\frac{1}{2}\right),
\end{equation}
where $E_\k=\sqrt{\Lambda_\k^2-\Delta_\k^2}\geq0$ is the triplon dispersion. The minimization of the ground state energy with respect to unknown mean-field parameters $\lambda$ and $\bar{s}^2$ leads to self-consistent equations which are solved as previously. 

% subsection plaquette_triplon_mean_field_theory (end)

% section triplon_mean_field_theory (end)

\section{Results} % (fold)
\label{sec:results}

For each set of varying exchange couplings of model~\eqref{eq:model_ss}, we find values of mean-field parameters of the triplon mean-field theory by solving self-consistent equations iteratively. Using mean-field parameters, we calculate physical observables like ground state energy, spin gap, staggered magnetization, etc. By knowing the observable values, the phase boundaries and nature of phase transitions are determined. In model Hamiltonian ~\eqref{eq:model_ss}, there are four exchange couplings, namely $J_1$, $J_2$, $J_3$, and $K$. If we vary them independently, we will find a four-dimensional phase diagram. This brute-force setup amplifies complexity and also poses difficulty in the result presentation. To avoid this, we shorten our task by considering three exchange couplings at a time. One can adiabatically connect the reduced phase diagram with the full phase diagram by choosing some fixed value of the remaining exchange coupling. We apply parametrization so that the three chosen exchange couplings always give a two-dimensional phase diagram. Below we present quantum phase diagrams in three different parameter spaces: $J_1$-$J_2$-$K$, $J_2$-$J_3$-$K$, and $J_1$-$J_3$-$K$.

\subsection{Quantum phase diagram in the $J_1$-$J_2$-$K$ space} % (fold)
\label{sub:quantum_phase_diagram_in_the_j_1_j_2_k_space}

\begin{figure}[htbp]
    \centering
        \includegraphics[width=.49\textwidth]{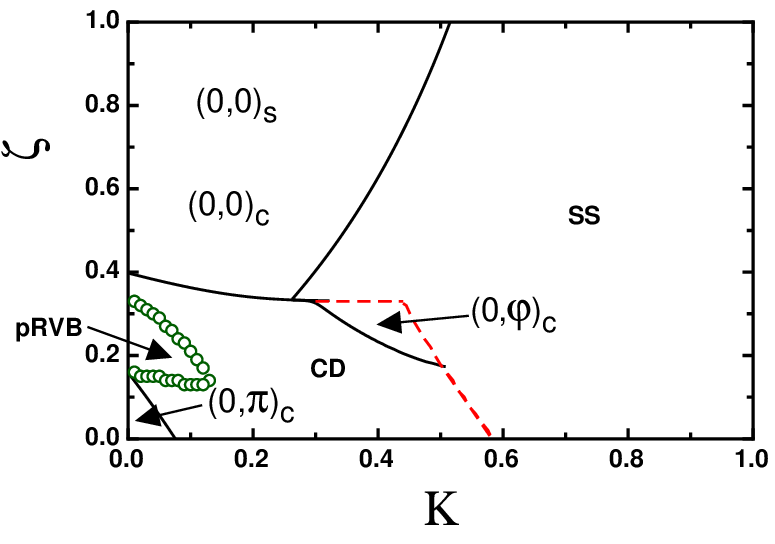}
    \caption{Quantum phase diagram in $J_1$-$J_2$-$K$ space. The solid and broken lines (or bubbles) denote continuous and level-crossing phase transitions, respectively.}
    \label{fig:QPD_J1J2}
\end{figure}

Here, the exchange couplings $J_1$ and $J_2$ are parametrized as $J_1=(1-\zeta)J$, $J_2=\zeta J$ such that $J+K=1$. As a result, we have left only two independent parameters $\zeta$ and $K$ which make possible to construct a two-dimensional phase diagram in the $\zeta$-$K$ space. Results from triplon mean-field theories with respect to SS, CD, and pRVB spin-gapped states are presented in the quantum phase diagram shown in Fig.~\ref{fig:QPD_J1J2}. 
Triplon mean-field theory with respect to SS state predicts wide regions of stable gapped SS state and a magnetically ordered collinear state $(0,0)_s$. But it does not give N\'eel state. This is not surprising as the $\mathcal{H}_K$ term does not contain any first neighbor interaction, and the $\mathcal{H}_{1}$ exchange interaction does not contribute in the triplon mean-field theory for a choice of SS dimerization. After incorporating the results of triplon mean-field theory for the underline CD state, the overestimated boundary of collinear state found earlier is shrunk. 
\begin{figure}[htbp]
  \centering
    \subfigure[$\k=(0,0)_s$]{\label{fig:j1j2_ss}\includegraphics[width=.21\textwidth]{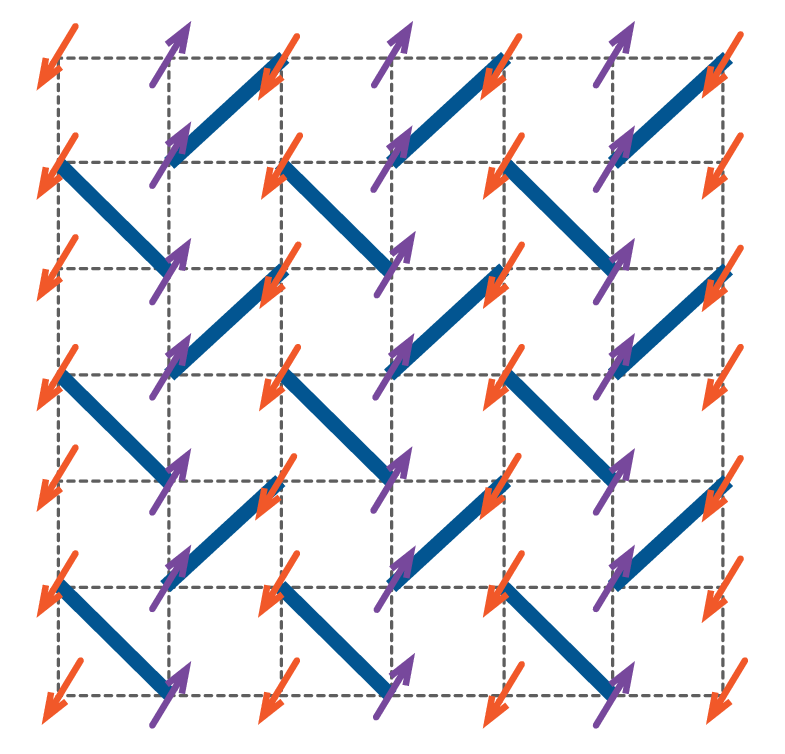}}
    \subfigure[$\k=(0,\pi)_c$]{\label{fig:neel}\includegraphics[width=.21\textwidth]{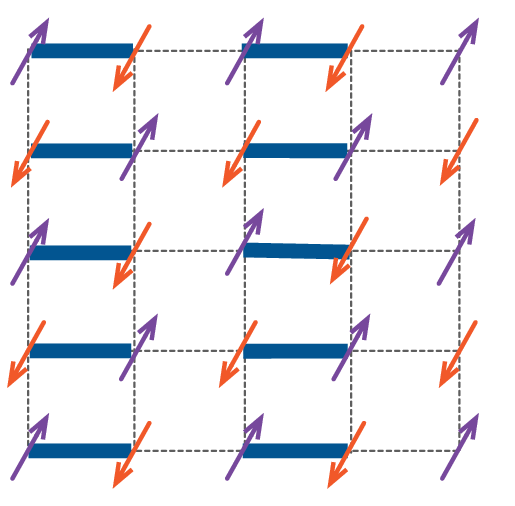}}
    \subfigure[$\k=(0,0)_c$]{\label{fig:collinear}\includegraphics[width=.21\textwidth]{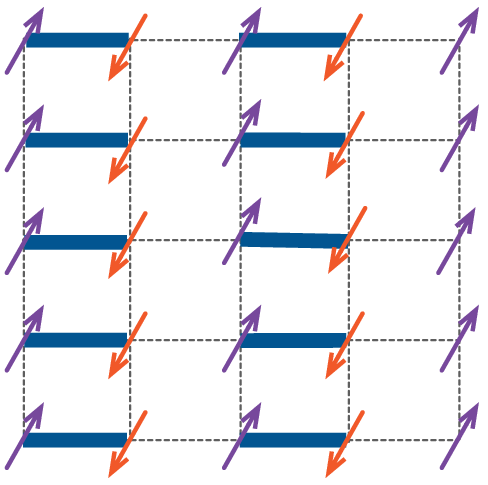}}
    \subfigure[$\k=(0,\varphi)_c$]{\label{fig:j1j2phi}\includegraphics[width=.20\textwidth]{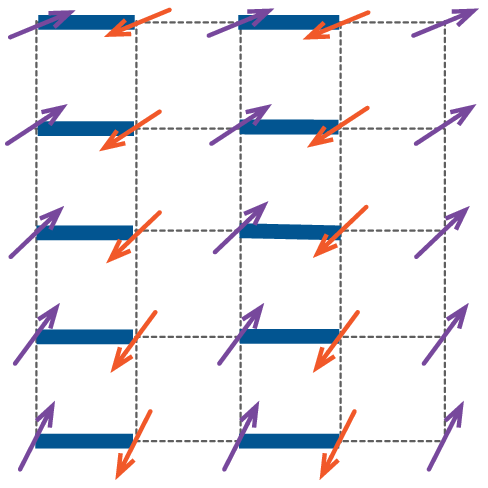}}
  \caption{Magnetic orderings of $(0,0)_s$, $(0,\pi)_c$, $(0,0)_c$ and $(0,\varphi)_c$ states. A subscript in a magnetically ordered state represents the underline dimer on top of which it develops. Here $s$ and $c$ respectively stand for SS and CD states.}
  \label{fig:j1j2kSSCD}
\end{figure}
The latter theory now correctly shows the N\'eel state $(0,\pi)_c$ in a small region. It also gives an incommensurate magnetically ordered state $(0,\varphi)_c$, where $\varphi=\arccos[(-J_1+2\tilde{J}_2)/4\tilde{J}_3]$. The stability of the pRVB state is checked from the results of plaquette triplon mean-field theory. It turns out that this state survives in a small window even in the presence of exchange coupling $K$. Recent experiments have confirmed the existence of pRVB state in strontium copper borate~\cite{Zayed2017} which can be mapped theoretically on the SS model~\cite{Kageyama1999,Shastry1981}. Our finding suggests that this state also exists in the degenerate SS model~\eqref{eq:model_ss}. Pictorial drawings of the magnetic orderings $(0,0)_s$, $(0,\pi)_c$, $(0,0)_c$ and $(0,\varphi)_c$ are shown in Fig.~\ref{fig:j1j2kSSCD}. The incommensurate magnetic order $(0,\varphi)_c$ shows a level-crossing phase transition with SS state, and a continuous transition with CD state. On the other hand, the collinear magnetic order, $(0,0)_s$ or $(0,0)_c$, undergoes a continuous phase transition with both SS and CD states. The antiferromagnetic N\'eel order also gives continuous phase transition with the CD state, which in turn displays level-crossing phase transition with SS and pRVB states. On the $\zeta=0$ axis in Fig.~\ref{fig:QPD_J1J2}, G\'elle \emph{et al.}~\cite{Gelle2008} have shown that the SS state continues to exist if the multi-spin exchange coupling satisfy the condition $K\gtrsim 0.6$. This is in good agreement with our mean-field results in which this condition comes out to be $K\gtrsim 0.58$. Moreover, the N\'eel state extends up to $K\lesssim 0.2$ on the $\zeta=0$ axis in exact diagonalization results and $K\lesssim 0.08$ in triplon mean-field theory. On the $\zeta=1$ axis, the exact diagonalization results show that the collinear state exists in the range $0\leq K \lesssim 0.87$, and our mean-field results say it to be $0\leq K \lesssim 0.52$. The underlying reason for these underestimations is that we always assume singlet condensation on the dimer or plaquette lattice units on top of which a magnetic order emerges. Hence, in general, an underestimated region of a magnetically ordered phase takes place in the triplon mean-field theory. This is confirmed by the $J_1$-$J_2$ Heisenberg antiferromagnet on square lattice in which the N\'eel state exits in the range $0\leq J_2/J_1\lesssim 0.4$~\cite{Chakravarty1989, Poilblanc1991, Schulz1996, Oitmaa1996, Capriotti2001, Richter2010} but the triplon mean-field theory suggests that this range is to be $0\leq J_2/J_1\lesssim 0.19$~\cite{Sachdev1990}. 

% subsection quantum_phase_diagram_in_the_j_1_j_2_k_space (end)

\subsection{Quantum phase diagram in the $J_2$-$J_3$-$K$ space} % (fold)
\label{sub:quantum_phase_diagram_in_the_j_3_j_3_k_space}

\begin{figure}[htbp]
  \centering
      \subfigure[$\k=(\pi/2,\pi/2)_s$ ]{\label{fig:j2j3_ss}\includegraphics[width=.18\textwidth]{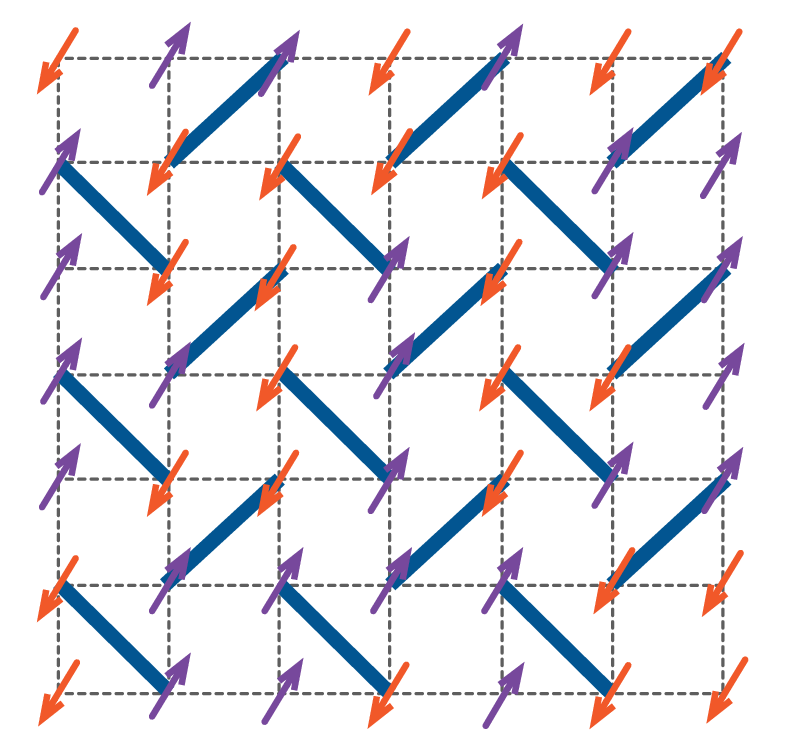}}
      \subfigure[$\k=(\pi/2,\varphi)_c$ ]{\label{fig:j1j3phi}\includegraphics[width=.28\textwidth]{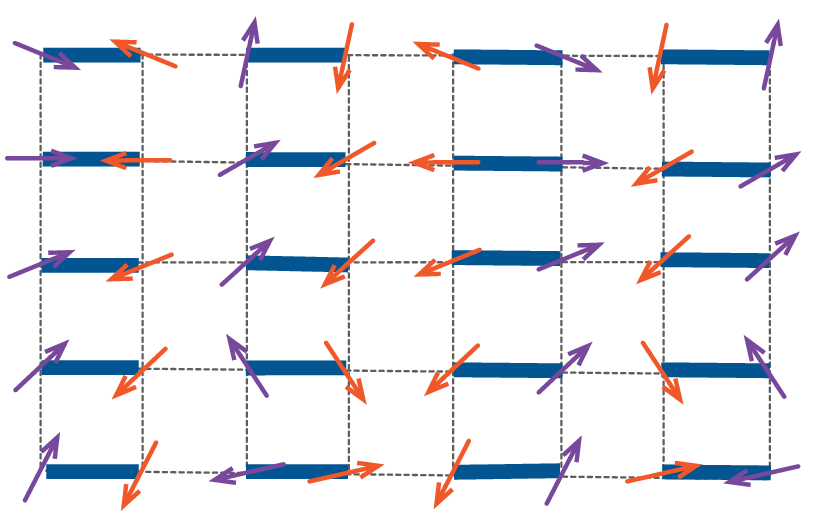}}
      \subfigure[$\k=(0,\pi)_{sc}$ ]{\label{fig:sCDS_0_pi}\includegraphics[width=.22\textwidth]{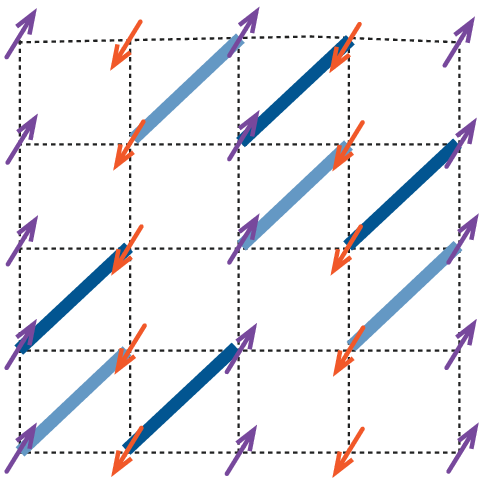}}
      \subfigure[$\k=(0,0)_{sc}$
      ]{\label{fig:sCDS_0_0}\includegraphics[width=.22\textwidth]{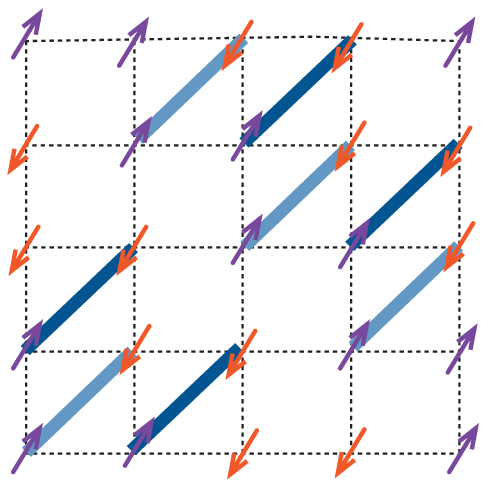}}
  \caption{Magnetically ordered phases with ordering wave vector $\k$. Here the subscript $sc$ denotes the underlying sCD state.}
  \label{fig:j2j3kSSCD}
\end{figure}

\begin{figure}[htbp]
    \centering
        \includegraphics[width=.49\textwidth]{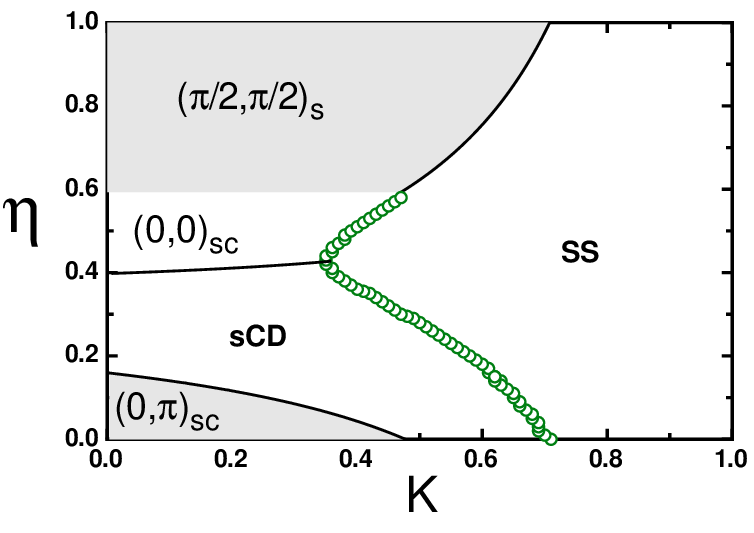}
    \caption{Quantum phase diagram in $J_2$-$J_3$-$K$ space.}
    \label{fig:QPD_J2J3}
\end{figure}

For 2D quantum phase diagram in the  $J_2$-$J_3$-$K$ space,  we parametrize $J_2$, $J_3$, and $K$ exchange-couplings as follows: $J_2=(1-\eta)J$ and $J_3=\eta J$ such that $J+K=1$. Here, the dimer triplon mean-field theory gives four magnetically ordered phases with wave vectors $(0,0)_{s}$, $(\pi/2,\pi/2)_{s}$,  $(0,0)_{sc}$, and $(0,\pi)_{sc}$ as pictorially drawn in Figs.~\ref{fig:j1j2kSSCD} and~\ref{fig:j2j3kSSCD}.
In the phase diagram, the phases $(0,0)_{s}$ and $(0,\pi)_{sc}$ coexist but the latter is found to be energetically more favorable than the previous one. Similarly, the $(0,0)_{sc}$ phase is energetically more favorable in a portion of $(\pi/2,\pi/2)_{s}$ phase. The resultant quantum phase is shown in Fig.~\ref{fig:QPD_J2J3} in which $(0,0)_{s}$ phase is not shown.
The phase $(0,0)_{sc}$ undergoes a continuous transition with sCD state, and level-crossing with SS state and $(\pi/2,\pi/2)_{s}$. Moreover, the magnetically ordered phases $(0,\pi)_{sc}$ and $(\pi/2,\pi/2)_{s}$ show continuous phase transitions with sCD and SS states, respectively. Between dimer states SS and sCD, a level-crossing phase transition is found. Let us now discuss the effect of some finite value of $J_1$ in the quantum phase diagram shown in Fig.~\ref{fig:QPD_J2J3}. The expression of $\Gamma$ in Eq.~\eqref{eq:gamma_scd} gives zero for both $(0,0)_{sc}$ and $(0,\pi)_{sc}$. Hence, these phases are stable against $J_1$. In SS dimer triplon mean-field theory in Sec.~\ref{sub:dimer_triplon_mean_field_theory}, there is no term containing $J_1$, and therefore the phases $(\pi/2,\pi/2)_{s}$ and SS are expected to be stable against any $J_1$ perturbation. Series expansion calculation of the SS model also says that the SS state does not depend on $J_1$~\cite{Koga2000}. 

% subsection quantum_phase_diagram_in_the_j_3_j_3_k_space (end)

\subsection{Quantum phase diagram in the $J_1$-$J_3$-$K$ space} % (fold)
\label{sub:quantum_phase_diagram_in_the_j_1_j_3_k_space}

\begin{figure}[htbp]
    \centering
    \includegraphics[width=0.49\textwidth]{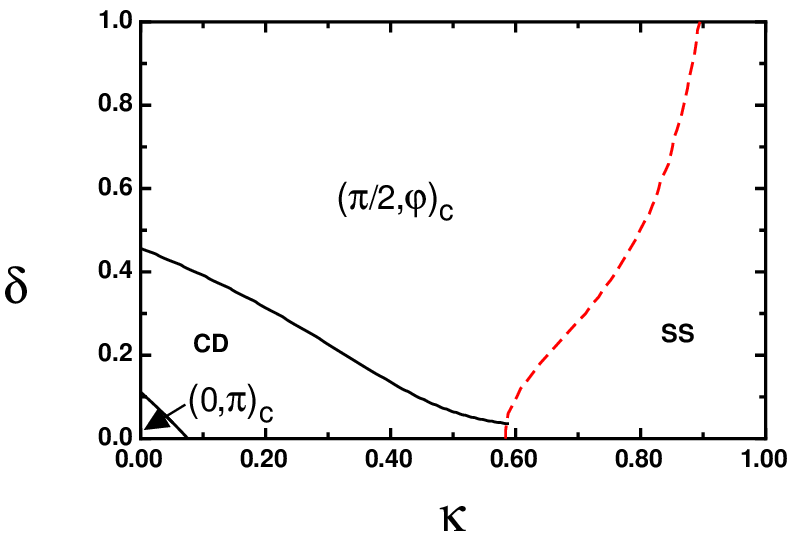}
    \caption{Quantum phase diagram in $J_1$-$J_3$-$K$ space.}
    \label{fig:QPD_J1J3}
\end{figure}

Here we follow the parametrization: $J_1=(1-\delta)J$, $J_3=\delta J$, where $J=1-K$. For these exchange coupling, the SS and CD states are the good choices of dimerization. Triplon mean-field theory of Hamiltonian~\eqref{eq:model_ss} for these dimer states gives a commensurate magnetic order $(0,\pi)_c$ and an incommensurate magnetic order $(\pi/2,\varphi)_c$ , where $\varphi=\arccos(-{J_1}/{4\tilde{J}_3})$. Pictorial drawing of these orders are shown in Figs.~\ref{fig:j1j2kSSCD} and~\ref{fig:j2j3kSSCD}. Complete quantum phase diagram is given in Fig.~\ref{fig:QPD_J1J3}.
Here both ordered phases $(0,\pi)_c$ and $(\pi/2,\varphi)_c$ undergo a continuous phase transition with $CD$ state, and the incommensurate magnetic order shows a level-crossing phase transition with the SS state. A level-crossing transition is found between the two dimer states CD and SS in a very narrow window.

% subsection quantum_phase_diagram_in_the_j_1_j_3_k_space (end)

% section results (end)

\section{Summary} % (fold)
\label{sec:summary}

We have performed calculation of triplon mean-field theory with respect to various dimer and plaquette states for an antiferromagnet with exact degenerate SS state. The model Hamiltonian of this antiferromagnet is an extension of the model proposed by Gell\'e \emph{et al.}~\cite{Gelle2008}, and the extension is done by adding an important third neighbor Heisenberg antiferromagnetic exchange interaction. Stabilities of the chosen spin-gapped states are searched by solving the self-consistent equations of the triplon mean-field theory. Recent experiments show that the plaquette state exits in the material SrCu$_2$(BO$_3$)$_2$ in which the SS model is realized~\cite{Zayed2017}. Remarkably,  our results also predict a stable plaquette state even though our considered model has degenerate SS states unlike the SS model. We also proposed an intra-sublattice dimer order sCD which gets stabilization by the third neighbor Heisenberg exchange interaction. We found that the quantum phase transition between two spin-gapped phases is always a level-crossing type phase transition whereas commensurate and incommensurate magnetic orders either undergo a level-crossing or a continuous phase transition.

% section summary (end)

\begin{acknowledgments}
R.K. thanks Pratyay Ghosh for some useful suggestions, and acknowledges JNU for the support through a Visiting Scholar position.
\end{acknowledgments}

\bibliography{ss2017}

\begin{thebibliography}{47}%
\makeatletter
\providecommand \@ifxundefined [1]{%
 \@ifx{#1\undefined}
}%
\providecommand \@ifnum [1]{%
 \ifnum #1\expandafter \@firstoftwo
 \else \expandafter \@secondoftwo
 \fi
}%
\providecommand \@ifx [1]{%
 \ifx #1\expandafter \@firstoftwo
 \else \expandafter \@secondoftwo
 \fi
}%
\providecommand \natexlab [1]{#1}%
\providecommand \enquote  [1]{``#1''}%
\providecommand \bibnamefont  [1]{#1}%
\providecommand \bibfnamefont [1]{#1}%
\providecommand \citenamefont [1]{#1}%
\providecommand \href@noop [0]{\@secondoftwo}%
\providecommand \href [0]{\begingroup \@sanitize@url \@href}%
\providecommand \@href[1]{\@@startlink{#1}\@@href}%
\providecommand \@@href[1]{\endgroup#1\@@endlink}%
\providecommand \@sanitize@url [0]{\catcode `\\12\catcode `\$12\catcode
  `\&12\catcode `\#12\catcode `\^12\catcode `\_12\catcode `\%12\relax}%
\providecommand \@@startlink[1]{}%
\providecommand \@@endlink[0]{}%
\providecommand \url  [0]{\begingroup\@sanitize@url \@url }%
\providecommand \@url [1]{\endgroup\@href {#1}{\urlprefix }}%
\providecommand \urlprefix  [0]{URL }%
\providecommand \Eprint [0]{\href }%
\providecommand \doibase [0]{http://dx.doi.org/}%
\providecommand \selectlanguage [0]{\@gobble}%
\providecommand \bibinfo  [0]{\@secondoftwo}%
\providecommand \bibfield  [0]{\@secondoftwo}%
\providecommand \translation [1]{[#1]}%
\providecommand \BibitemOpen [0]{}%
\providecommand \bibitemStop [0]{}%
\providecommand \bibitemNoStop [0]{.\EOS\space}%
\providecommand \EOS [0]{\spacefactor3000\relax}%
\providecommand \BibitemShut  [1]{\csname bibitem#1\endcsname}%
\let\auto@bib@innerbib\@empty
%</preamble>
\bibitem [{\citenamefont {Lacroix}\ \emph {et~al.}(2011)\citenamefont
  {Lacroix}, \citenamefont {Mendels},\ and\ \citenamefont
  {Mila}}]{Lacroix2011}%
  \BibitemOpen
  \bibinfo {editor} {\bibfnamefont {C.}~\bibnamefont {Lacroix}}, \bibinfo
  {editor} {\bibfnamefont {P.}~\bibnamefont {Mendels}}, \ and\ \bibinfo
  {editor} {\bibfnamefont {F.}~\bibnamefont {Mila}},\ eds.,\ \href {\doibase
  10.1007/978-3-642-10589-0} {\emph {\bibinfo {title} {Introduction to
  Frustrated Magnetism}}}\ (\bibinfo  {publisher} {Springer Berlin
  Heidelberg},\ \bibinfo {year} {2011})\BibitemShut {NoStop}%
\bibitem [{\citenamefont {Balents}(2010)}]{Balents2010}%
  \BibitemOpen
  \bibfield  {author} {\bibinfo {author} {\bibfnamefont {L.}~\bibnamefont
  {Balents}},\ }\href {http://dx.doi.org/10.1038/nature08917} {\bibfield
  {journal} {\bibinfo  {journal} {Nature}\ }\textbf {\bibinfo {volume} {464}},\
  \bibinfo {pages} {199} (\bibinfo {year} {2010})}\BibitemShut {NoStop}%
\bibitem [{\citenamefont {Chen}\ \emph {et~al.}(2012)\citenamefont {Chen},
  \citenamefont {Tao}, \citenamefont {Yao},\ and\ \citenamefont
  {Liu}}]{Chen2012}%
  \BibitemOpen
  \bibfield  {author} {\bibinfo {author} {\bibfnamefont {Y.-H.}\ \bibnamefont
  {Chen}}, \bibinfo {author} {\bibfnamefont {H.-S.}\ \bibnamefont {Tao}},
  \bibinfo {author} {\bibfnamefont {D.-X.}\ \bibnamefont {Yao}}, \ and\
  \bibinfo {author} {\bibfnamefont {W.-M.}\ \bibnamefont {Liu}},\ }\href
  {\doibase 10.1103/PhysRevLett.108.246402} {\bibfield  {journal} {\bibinfo
  {journal} {Phys. Rev. Lett.}\ }\textbf {\bibinfo {volume} {108}},\ \bibinfo
  {pages} {246402} (\bibinfo {year} {2012})}\BibitemShut {NoStop}%
\bibitem [{\citenamefont {Zhang}\ \emph {et~al.}(2009)\citenamefont {Zhang},
  \citenamefont {Hu}, \citenamefont {Bernevig}, \citenamefont {Wang},
  \citenamefont {Xie},\ and\ \citenamefont {Liu}}]{Zhang2009}%
  \BibitemOpen
  \bibfield  {author} {\bibinfo {author} {\bibfnamefont {Y.-Y.}\ \bibnamefont
  {Zhang}}, \bibinfo {author} {\bibfnamefont {J.}~\bibnamefont {Hu}}, \bibinfo
  {author} {\bibfnamefont {B.~A.}\ \bibnamefont {Bernevig}}, \bibinfo {author}
  {\bibfnamefont {X.~R.}\ \bibnamefont {Wang}}, \bibinfo {author}
  {\bibfnamefont {X.~C.}\ \bibnamefont {Xie}}, \ and\ \bibinfo {author}
  {\bibfnamefont {W.~M.}\ \bibnamefont {Liu}},\ }\href {\doibase
  10.1103/PhysRevLett.102.106401} {\bibfield  {journal} {\bibinfo  {journal}
  {Phys. Rev. Lett.}\ }\textbf {\bibinfo {volume} {102}},\ \bibinfo {pages}
  {106401} (\bibinfo {year} {2009})}\BibitemShut {NoStop}%
\bibitem [{\citenamefont {Ji}\ \emph {et~al.}(2007)\citenamefont {Ji},
  \citenamefont {Xie},\ and\ \citenamefont {Liu}}]{Ji2007}%
  \BibitemOpen
  \bibfield  {author} {\bibinfo {author} {\bibfnamefont {A.-C.}\ \bibnamefont
  {Ji}}, \bibinfo {author} {\bibfnamefont {X.~C.}\ \bibnamefont {Xie}}, \ and\
  \bibinfo {author} {\bibfnamefont {W.~M.}\ \bibnamefont {Liu}},\ }\href
  {\doibase 10.1103/PhysRevLett.99.183602} {\bibfield  {journal} {\bibinfo
  {journal} {Phys. Rev. Lett.}\ }\textbf {\bibinfo {volume} {99}},\ \bibinfo
  {pages} {183602} (\bibinfo {year} {2007})}\BibitemShut {NoStop}%
\bibitem [{\citenamefont {Misguich}\ and\ \citenamefont
  {Lhuillier}(2013)}]{Misguich2013}%
  \BibitemOpen
  \bibfield  {author} {\bibinfo {author} {\bibfnamefont {G.}~\bibnamefont
  {Misguich}}\ and\ \bibinfo {author} {\bibfnamefont {C.}~\bibnamefont
  {Lhuillier}},\ }in\ \href {\doibase 10.1142/9789814440745_0005} {\emph
  {\bibinfo {booktitle} {Frustrated Spin Systems}}}\ (\bibinfo  {publisher}
  {{World} {Scientific}},\ \bibinfo {year} {2013})\ pp.\ \bibinfo {pages}
  {235--319}\BibitemShut {NoStop}%
\bibitem [{\citenamefont {Miyahara}\ and\ \citenamefont
  {Ueda}(2003)}]{Miyahara2003}%
  \BibitemOpen
  \bibfield  {author} {\bibinfo {author} {\bibfnamefont {S.}~\bibnamefont
  {Miyahara}}\ and\ \bibinfo {author} {\bibfnamefont {K.}~\bibnamefont
  {Ueda}},\ }\href {http://stacks.iop.org/0953-8984/15/i=9/a=201} {\bibfield
  {journal} {\bibinfo  {journal} {Journal of Physics: Condensed Matter}\
  }\textbf {\bibinfo {volume} {15}},\ \bibinfo {pages} {R327} (\bibinfo {year}
  {2003})}\BibitemShut {NoStop}%
\bibitem [{\citenamefont {Majumdar}\ and\ \citenamefont
  {Ghosh}(1969{\natexlab{a}})}]{Majumdar1969}%
  \BibitemOpen
  \bibfield  {author} {\bibinfo {author} {\bibfnamefont {C.~K.}\ \bibnamefont
  {Majumdar}}\ and\ \bibinfo {author} {\bibfnamefont {D.~K.}\ \bibnamefont
  {Ghosh}},\ }\href {\doibase 10.1063/1.1664978} {\bibfield  {journal}
  {\bibinfo  {journal} {Journal of Mathematical Physics}\ }\textbf {\bibinfo
  {volume} {10}},\ \bibinfo {pages} {1388} (\bibinfo {year}
  {1969}{\natexlab{a}})}\BibitemShut {NoStop}%
\bibitem [{\citenamefont {Majumdar}\ and\ \citenamefont
  {Ghosh}(1969{\natexlab{b}})}]{Majumdar1969a}%
  \BibitemOpen
  \bibfield  {author} {\bibinfo {author} {\bibfnamefont {C.~K.}\ \bibnamefont
  {Majumdar}}\ and\ \bibinfo {author} {\bibfnamefont {D.~K.}\ \bibnamefont
  {Ghosh}},\ }\href {\doibase 10.1063/1.1664979} {\bibfield  {journal}
  {\bibinfo  {journal} {Journal of Mathematical Physics}\ }\textbf {\bibinfo
  {volume} {10}},\ \bibinfo {pages} {1399} (\bibinfo {year}
  {1969}{\natexlab{b}})}\BibitemShut {NoStop}%
\bibitem [{\citenamefont {Hase}\ \emph {et~al.}(1993)\citenamefont {Hase},
  \citenamefont {Terasaki},\ and\ \citenamefont {Uchinokura}}]{Hase1993}%
  \BibitemOpen
  \bibfield  {author} {\bibinfo {author} {\bibfnamefont {M.}~\bibnamefont
  {Hase}}, \bibinfo {author} {\bibfnamefont {I.}~\bibnamefont {Terasaki}}, \
  and\ \bibinfo {author} {\bibfnamefont {K.}~\bibnamefont {Uchinokura}},\
  }\href {\doibase 10.1103/PhysRevLett.70.3651} {\bibfield  {journal} {\bibinfo
   {journal} {Phys. Rev. Lett.}\ }\textbf {\bibinfo {volume} {70}},\ \bibinfo
  {pages} {3651} (\bibinfo {year} {1993})}\BibitemShut {NoStop}%
\bibitem [{\citenamefont {Castilla}\ \emph {et~al.}(1995)\citenamefont
  {Castilla}, \citenamefont {Chakravarty},\ and\ \citenamefont
  {Emery}}]{Castilla1995}%
  \BibitemOpen
  \bibfield  {author} {\bibinfo {author} {\bibfnamefont {G.}~\bibnamefont
  {Castilla}}, \bibinfo {author} {\bibfnamefont {S.}~\bibnamefont
  {Chakravarty}}, \ and\ \bibinfo {author} {\bibfnamefont {V.~J.}\ \bibnamefont
  {Emery}},\ }\href {\doibase 10.1103/PhysRevLett.75.1823} {\bibfield
  {journal} {\bibinfo  {journal} {Phys. Rev. Lett.}\ }\textbf {\bibinfo
  {volume} {75}},\ \bibinfo {pages} {1823} (\bibinfo {year}
  {1995})}\BibitemShut {NoStop}%
\bibitem [{\citenamefont {Shastry}\ and\ \citenamefont
  {Sutherland}(1981)}]{Shastry1981}%
  \BibitemOpen
  \bibfield  {author} {\bibinfo {author} {\bibfnamefont {B.~S.}\ \bibnamefont
  {Shastry}}\ and\ \bibinfo {author} {\bibfnamefont {B.}~\bibnamefont
  {Sutherland}},\ }\href {\doibase 10.1016/0378-4363(81)90838-x} {\bibfield
  {journal} {\bibinfo  {journal} {Physica B+C}\ }\textbf {\bibinfo {volume}
  {108}},\ \bibinfo {pages} {1069} (\bibinfo {year} {1981})}\BibitemShut
  {NoStop}%
\bibitem [{\citenamefont {Kageyama}\ \emph {et~al.}(1999)\citenamefont
  {Kageyama}, \citenamefont {Yoshimura}, \citenamefont {Stern}, \citenamefont
  {Mushnikov}, \citenamefont {Onizuka}, \citenamefont {Kato}, \citenamefont
  {Kosuge}, \citenamefont {Slichter}, \citenamefont {Goto},\ and\ \citenamefont
  {Ueda}}]{Kageyama1999}%
  \BibitemOpen
  \bibfield  {author} {\bibinfo {author} {\bibfnamefont {H.}~\bibnamefont
  {Kageyama}}, \bibinfo {author} {\bibfnamefont {K.}~\bibnamefont {Yoshimura}},
  \bibinfo {author} {\bibfnamefont {R.}~\bibnamefont {Stern}}, \bibinfo
  {author} {\bibfnamefont {N.~V.}\ \bibnamefont {Mushnikov}}, \bibinfo {author}
  {\bibfnamefont {K.}~\bibnamefont {Onizuka}}, \bibinfo {author} {\bibfnamefont
  {M.}~\bibnamefont {Kato}}, \bibinfo {author} {\bibfnamefont {K.}~\bibnamefont
  {Kosuge}}, \bibinfo {author} {\bibfnamefont {C.~P.}\ \bibnamefont
  {Slichter}}, \bibinfo {author} {\bibfnamefont {T.}~\bibnamefont {Goto}}, \
  and\ \bibinfo {author} {\bibfnamefont {Y.}~\bibnamefont {Ueda}},\ }\href
  {\doibase 10.1103/PhysRevLett.82.3168} {\bibfield  {journal} {\bibinfo
  {journal} {Phys. Rev. Lett.}\ }\textbf {\bibinfo {volume} {82}},\ \bibinfo
  {pages} {3168} (\bibinfo {year} {1999})}\BibitemShut {NoStop}%
\bibitem [{\citenamefont {Miyahara}\ and\ \citenamefont
  {Ueda}(1999)}]{Miyahara1999}%
  \BibitemOpen
  \bibfield  {author} {\bibinfo {author} {\bibfnamefont {S.}~\bibnamefont
  {Miyahara}}\ and\ \bibinfo {author} {\bibfnamefont {K.}~\bibnamefont
  {Ueda}},\ }\href {\doibase 10.1103/PhysRevLett.82.3701} {\bibfield  {journal}
  {\bibinfo  {journal} {Phys. Rev. Lett.}\ }\textbf {\bibinfo {volume} {82}},\
  \bibinfo {pages} {3701} (\bibinfo {year} {1999})}\BibitemShut {NoStop}%
\bibitem [{\citenamefont {Koga}\ and\ \citenamefont
  {Kawakami}(2000)}]{Koga2000}%
  \BibitemOpen
  \bibfield  {author} {\bibinfo {author} {\bibfnamefont {A.}~\bibnamefont
  {Koga}}\ and\ \bibinfo {author} {\bibfnamefont {N.}~\bibnamefont
  {Kawakami}},\ }\href {\doibase 10.1103/PhysRevLett.84.4461} {\bibfield
  {journal} {\bibinfo  {journal} {Phys. Rev. Lett.}\ }\textbf {\bibinfo
  {volume} {84}},\ \bibinfo {pages} {4461} (\bibinfo {year}
  {2000})}\BibitemShut {NoStop}%
\bibitem [{\citenamefont {L\"auchli}\ \emph {et~al.}(2002)\citenamefont
  {L\"auchli}, \citenamefont {Wessel},\ and\ \citenamefont
  {Sigrist}}]{Lauchli2002}%
  \BibitemOpen
  \bibfield  {author} {\bibinfo {author} {\bibfnamefont {A.}~\bibnamefont
  {L\"auchli}}, \bibinfo {author} {\bibfnamefont {S.}~\bibnamefont {Wessel}}, \
  and\ \bibinfo {author} {\bibfnamefont {M.}~\bibnamefont {Sigrist}},\ }\href
  {\doibase 10.1103/PhysRevB.66.014401} {\bibfield  {journal} {\bibinfo
  {journal} {Phys. Rev. B}\ }\textbf {\bibinfo {volume} {66}},\ \bibinfo
  {pages} {014401} (\bibinfo {year} {2002})}\BibitemShut {NoStop}%
\bibitem [{\citenamefont {Zhang}\ and\ \citenamefont
  {Sengupta}(2015)}]{Zhang2015}%
  \BibitemOpen
  \bibfield  {author} {\bibinfo {author} {\bibfnamefont {Z.}~\bibnamefont
  {Zhang}}\ and\ \bibinfo {author} {\bibfnamefont {P.}~\bibnamefont
  {Sengupta}},\ }\href {\doibase 10.1103/PhysRevB.92.094440} {\bibfield
  {journal} {\bibinfo  {journal} {Phys. Rev. B}\ }\textbf {\bibinfo {volume}
  {92}},\ \bibinfo {pages} {094440} (\bibinfo {year} {2015})}\BibitemShut
  {NoStop}%
\bibitem [{\citenamefont {Zayed}\ \emph {et~al.}(2017)\citenamefont {Zayed},
  \citenamefont {R{\"u}egg}, \citenamefont {Larrea~J.}, \citenamefont
  {L{\"a}uchli}, \citenamefont {Panagopoulos}, \citenamefont {Saxena},
  \citenamefont {Ellerby}, \citenamefont {McMorrow}, \citenamefont
  {Str{\"a}ssle}, \citenamefont {Klotz}, \citenamefont {Hamel}, \citenamefont
  {Sadykov}, \citenamefont {Pomjakushin}, \citenamefont {Boehm}, \citenamefont
  {Jim{\'e}nez-Ruiz}, \citenamefont {Schneidewind}, \citenamefont
  {Pomjakushina}, \citenamefont {Stingaciu}, \citenamefont {Conder},\ and\
  \citenamefont {R{\o}nnow}}]{Zayed2017}%
  \BibitemOpen
  \bibfield  {author} {\bibinfo {author} {\bibfnamefont {M.~E.}\ \bibnamefont
  {Zayed}}, \bibinfo {author} {\bibfnamefont {C.}~\bibnamefont {R{\"u}egg}},
  \bibinfo {author} {\bibfnamefont {J.}~\bibnamefont {Larrea~J.}}, \bibinfo
  {author} {\bibfnamefont {A.~M.}\ \bibnamefont {L{\"a}uchli}}, \bibinfo
  {author} {\bibfnamefont {C.}~\bibnamefont {Panagopoulos}}, \bibinfo {author}
  {\bibfnamefont {S.~S.}\ \bibnamefont {Saxena}}, \bibinfo {author}
  {\bibfnamefont {M.}~\bibnamefont {Ellerby}}, \bibinfo {author} {\bibfnamefont
  {D.~F.}\ \bibnamefont {McMorrow}}, \bibinfo {author} {\bibfnamefont
  {T.}~\bibnamefont {Str{\"a}ssle}}, \bibinfo {author} {\bibfnamefont
  {S.}~\bibnamefont {Klotz}}, \bibinfo {author} {\bibfnamefont
  {G.}~\bibnamefont {Hamel}}, \bibinfo {author} {\bibfnamefont {R.~A.}\
  \bibnamefont {Sadykov}}, \bibinfo {author} {\bibfnamefont {V.}~\bibnamefont
  {Pomjakushin}}, \bibinfo {author} {\bibfnamefont {M.}~\bibnamefont {Boehm}},
  \bibinfo {author} {\bibfnamefont {M.}~\bibnamefont {Jim{\'e}nez-Ruiz}},
  \bibinfo {author} {\bibfnamefont {A.}~\bibnamefont {Schneidewind}}, \bibinfo
  {author} {\bibfnamefont {E.}~\bibnamefont {Pomjakushina}}, \bibinfo {author}
  {\bibfnamefont {M.}~\bibnamefont {Stingaciu}}, \bibinfo {author}
  {\bibfnamefont {K.}~\bibnamefont {Conder}}, \ and\ \bibinfo {author}
  {\bibfnamefont {H.~M.}\ \bibnamefont {R{\o}nnow}},\ }\href
  {http://dx.doi.org/10.1038/nphys4190} {\bibfield  {journal} {\bibinfo
  {journal} {Nature Physics}\ }\textbf {\bibinfo {volume} {13}},\ \bibinfo
  {pages} {962 EP } (\bibinfo {year} {2017})}\BibitemShut {NoStop}%
\bibitem [{\citenamefont {Kumar}(2002)}]{Kumar2002}%
  \BibitemOpen
  \bibfield  {author} {\bibinfo {author} {\bibfnamefont {B.}~\bibnamefont
  {Kumar}},\ }\href {\doibase 10.1103/PhysRevB.66.024406} {\bibfield  {journal}
  {\bibinfo  {journal} {Phys. Rev. B}\ }\textbf {\bibinfo {volume} {66}},\
  \bibinfo {pages} {024406} (\bibinfo {year} {2002})}\BibitemShut {NoStop}%
\bibitem [{\citenamefont {Surendran}\ and\ \citenamefont
  {Shankar}(2002)}]{Surendran2002}%
  \BibitemOpen
  \bibfield  {author} {\bibinfo {author} {\bibfnamefont {N.}~\bibnamefont
  {Surendran}}\ and\ \bibinfo {author} {\bibfnamefont {R.}~\bibnamefont
  {Shankar}},\ }\href {\doibase 10.1103/PhysRevB.66.024415} {\bibfield
  {journal} {\bibinfo  {journal} {Phys. Rev. B}\ }\textbf {\bibinfo {volume}
  {66}},\ \bibinfo {pages} {024415} (\bibinfo {year} {2002})}\BibitemShut
  {NoStop}%
\bibitem [{\citenamefont {Schmidt}(2005)}]{Schmidt2005}%
  \BibitemOpen
  \bibfield  {author} {\bibinfo {author} {\bibfnamefont {H.-J.}\ \bibnamefont
  {Schmidt}},\ }\href {\doibase 10.1088/0305-4470/38/10/005} {\bibfield
  {journal} {\bibinfo  {journal} {Journal of Physics A: Mathematical and
  General}\ }\textbf {\bibinfo {volume} {38}},\ \bibinfo {pages} {2123}
  (\bibinfo {year} {2005})}\BibitemShut {NoStop}%
\bibitem [{\citenamefont {Danu}\ \emph {et~al.}(2012)\citenamefont {Danu},
  \citenamefont {Kumar},\ and\ \citenamefont {Pai}}]{Danu2012}%
  \BibitemOpen
  \bibfield  {author} {\bibinfo {author} {\bibfnamefont {B.}~\bibnamefont
  {Danu}}, \bibinfo {author} {\bibfnamefont {B.}~\bibnamefont {Kumar}}, \ and\
  \bibinfo {author} {\bibfnamefont {R.~V.}\ \bibnamefont {Pai}},\ }\href
  {\doibase 10.1209/0295-5075/100/27003} {\bibfield  {journal} {\bibinfo
  {journal} {{EPL} (Europhysics Letters)}\ }\textbf {\bibinfo {volume} {100}},\
  \bibinfo {pages} {27003} (\bibinfo {year} {2012})}\BibitemShut {NoStop}%
\bibitem [{\citenamefont {Kumar}\ and\ \citenamefont
  {Kumar}(2008)}]{Kumar2008}%
  \BibitemOpen
  \bibfield  {author} {\bibinfo {author} {\bibfnamefont {R.}~\bibnamefont
  {Kumar}}\ and\ \bibinfo {author} {\bibfnamefont {B.}~\bibnamefont {Kumar}},\
  }\href {\doibase 10.1103/PhysRevB.77.144413} {\bibfield  {journal} {\bibinfo
  {journal} {Phys. Rev. B}\ }\textbf {\bibinfo {volume} {77}},\ \bibinfo
  {pages} {144413} (\bibinfo {year} {2008})}\BibitemShut {NoStop}%
\bibitem [{\citenamefont {Kumar}\ \emph {et~al.}(2009)\citenamefont {Kumar},
  \citenamefont {Kumar},\ and\ \citenamefont {Kumar}}]{Kumar2009}%
  \BibitemOpen
  \bibfield  {author} {\bibinfo {author} {\bibfnamefont {R.}~\bibnamefont
  {Kumar}}, \bibinfo {author} {\bibfnamefont {D.}~\bibnamefont {Kumar}}, \ and\
  \bibinfo {author} {\bibfnamefont {B.}~\bibnamefont {Kumar}},\ }\href
  {\doibase 10.1103/PhysRevB.80.214428} {\bibfield  {journal} {\bibinfo
  {journal} {Phys. Rev. B}\ }\textbf {\bibinfo {volume} {80}},\ \bibinfo
  {pages} {214428} (\bibinfo {year} {2009})}\BibitemShut {NoStop}%
\bibitem [{\citenamefont {Chakravarty}\ \emph {et~al.}(1989)\citenamefont
  {Chakravarty}, \citenamefont {Halperin},\ and\ \citenamefont
  {Nelson}}]{Chakravarty1989}%
  \BibitemOpen
  \bibfield  {author} {\bibinfo {author} {\bibfnamefont {S.}~\bibnamefont
  {Chakravarty}}, \bibinfo {author} {\bibfnamefont {B.~I.}\ \bibnamefont
  {Halperin}}, \ and\ \bibinfo {author} {\bibfnamefont {D.~R.}\ \bibnamefont
  {Nelson}},\ }\href {\doibase 10.1103/PhysRevB.39.2344} {\bibfield  {journal}
  {\bibinfo  {journal} {Phys. Rev. B}\ }\textbf {\bibinfo {volume} {39}},\
  \bibinfo {pages} {2344} (\bibinfo {year} {1989})}\BibitemShut {NoStop}%
\bibitem [{\citenamefont {Sachdev}\ and\ \citenamefont
  {Bhatt}(1990)}]{Sachdev1990}%
  \BibitemOpen
  \bibfield  {author} {\bibinfo {author} {\bibfnamefont {S.}~\bibnamefont
  {Sachdev}}\ and\ \bibinfo {author} {\bibfnamefont {R.~N.}\ \bibnamefont
  {Bhatt}},\ }\href {\doibase 10.1103/PhysRevB.41.9323} {\bibfield  {journal}
  {\bibinfo  {journal} {Phys. Rev. B}\ }\textbf {\bibinfo {volume} {41}},\
  \bibinfo {pages} {9323} (\bibinfo {year} {1990})}\BibitemShut {NoStop}%
\bibitem [{\citenamefont {Poilblanc}\ \emph {et~al.}(1991)\citenamefont
  {Poilblanc}, \citenamefont {Gagliano}, \citenamefont {Bacci},\ and\
  \citenamefont {Dagotto}}]{Poilblanc1991}%
  \BibitemOpen
  \bibfield  {author} {\bibinfo {author} {\bibfnamefont {D.}~\bibnamefont
  {Poilblanc}}, \bibinfo {author} {\bibfnamefont {E.}~\bibnamefont {Gagliano}},
  \bibinfo {author} {\bibfnamefont {S.}~\bibnamefont {Bacci}}, \ and\ \bibinfo
  {author} {\bibfnamefont {E.}~\bibnamefont {Dagotto}},\ }\href {\doibase
  10.1103/PhysRevB.43.10970} {\bibfield  {journal} {\bibinfo  {journal} {Phys.
  Rev. B}\ }\textbf {\bibinfo {volume} {43}},\ \bibinfo {pages} {10970}
  (\bibinfo {year} {1991})}\BibitemShut {NoStop}%
\bibitem [{\citenamefont {Schulz}\ \emph {et~al.}(1996)\citenamefont {Schulz},
  \citenamefont {Ziman},\ and\ \citenamefont {Poilblanc}}]{Schulz1996}%
  \BibitemOpen
  \bibfield  {author} {\bibinfo {author} {\bibfnamefont {H.~J.}\ \bibnamefont
  {Schulz}}, \bibinfo {author} {\bibfnamefont {T.~A.~L.}\ \bibnamefont
  {Ziman}}, \ and\ \bibinfo {author} {\bibfnamefont {D.}~\bibnamefont
  {Poilblanc}},\ }\href {http://dx.doi.org/10.1051/jp1:1996236} {\bibfield
  {journal} {\bibinfo  {journal} {J. Phys. I France}\ }\textbf {\bibinfo
  {volume} {6}},\ \bibinfo {pages} {675} (\bibinfo {year} {1996})}\BibitemShut
  {NoStop}%
\bibitem [{\citenamefont {Oitmaa}\ and\ \citenamefont
  {Weihong}(1996)}]{Oitmaa1996}%
  \BibitemOpen
  \bibfield  {author} {\bibinfo {author} {\bibfnamefont {J.}~\bibnamefont
  {Oitmaa}}\ and\ \bibinfo {author} {\bibfnamefont {Z.}~\bibnamefont
  {Weihong}},\ }\href {\doibase 10.1103/PhysRevB.54.3022} {\bibfield  {journal}
  {\bibinfo  {journal} {Phys. Rev. B}\ }\textbf {\bibinfo {volume} {54}},\
  \bibinfo {pages} {3022} (\bibinfo {year} {1996})}\BibitemShut {NoStop}%
\bibitem [{\citenamefont {Singh}\ \emph {et~al.}(1999)\citenamefont {Singh},
  \citenamefont {Weihong}, \citenamefont {Hamer},\ and\ \citenamefont
  {Oitmaa}}]{Singh1999}%
  \BibitemOpen
  \bibfield  {author} {\bibinfo {author} {\bibfnamefont {R.~R.~P.}\
  \bibnamefont {Singh}}, \bibinfo {author} {\bibfnamefont {Z.}~\bibnamefont
  {Weihong}}, \bibinfo {author} {\bibfnamefont {C.~J.}\ \bibnamefont {Hamer}},
  \ and\ \bibinfo {author} {\bibfnamefont {J.}~\bibnamefont {Oitmaa}},\ }\href
  {\doibase 10.1103/PhysRevB.60.7278} {\bibfield  {journal} {\bibinfo
  {journal} {Phys. Rev. B}\ }\textbf {\bibinfo {volume} {60}},\ \bibinfo
  {pages} {7278} (\bibinfo {year} {1999})}\BibitemShut {NoStop}%
\bibitem [{\citenamefont {Capriotti}\ \emph {et~al.}(2001)\citenamefont
  {Capriotti}, \citenamefont {Becca}, \citenamefont {Parola},\ and\
  \citenamefont {Sorella}}]{Capriotti2001}%
  \BibitemOpen
  \bibfield  {author} {\bibinfo {author} {\bibfnamefont {L.}~\bibnamefont
  {Capriotti}}, \bibinfo {author} {\bibfnamefont {F.}~\bibnamefont {Becca}},
  \bibinfo {author} {\bibfnamefont {A.}~\bibnamefont {Parola}}, \ and\ \bibinfo
  {author} {\bibfnamefont {S.}~\bibnamefont {Sorella}},\ }\href {\doibase
  10.1103/PhysRevLett.87.097201} {\bibfield  {journal} {\bibinfo  {journal}
  {Phys. Rev. Lett.}\ }\textbf {\bibinfo {volume} {87}},\ \bibinfo {pages}
  {097201} (\bibinfo {year} {2001})}\BibitemShut {NoStop}%
\bibitem [{\citenamefont {Sushkov}\ \emph {et~al.}(2001)\citenamefont
  {Sushkov}, \citenamefont {Oitmaa},\ and\ \citenamefont
  {Weihong}}]{Sushkov2001}%
  \BibitemOpen
  \bibfield  {author} {\bibinfo {author} {\bibfnamefont {O.~P.}\ \bibnamefont
  {Sushkov}}, \bibinfo {author} {\bibfnamefont {J.}~\bibnamefont {Oitmaa}}, \
  and\ \bibinfo {author} {\bibfnamefont {Z.}~\bibnamefont {Weihong}},\ }\href
  {\doibase 10.1103/PhysRevB.63.104420} {\bibfield  {journal} {\bibinfo
  {journal} {Phys. Rev. B}\ }\textbf {\bibinfo {volume} {63}},\ \bibinfo
  {pages} {104420} (\bibinfo {year} {2001})}\BibitemShut {NoStop}%
\bibitem [{\citenamefont {Richter}\ \emph {et~al.}(2010)\citenamefont
  {Richter}, \citenamefont {Darradi}, \citenamefont {Schulenburg},
  \citenamefont {Farnell},\ and\ \citenamefont {Rosner}}]{Richter2010}%
  \BibitemOpen
  \bibfield  {author} {\bibinfo {author} {\bibfnamefont {J.}~\bibnamefont
  {Richter}}, \bibinfo {author} {\bibfnamefont {R.}~\bibnamefont {Darradi}},
  \bibinfo {author} {\bibfnamefont {J.}~\bibnamefont {Schulenburg}}, \bibinfo
  {author} {\bibfnamefont {D.~J.~J.}\ \bibnamefont {Farnell}}, \ and\ \bibinfo
  {author} {\bibfnamefont {H.}~\bibnamefont {Rosner}},\ }\href {\doibase
  10.1103/PhysRevB.81.174429} {\bibfield  {journal} {\bibinfo  {journal} {Phys.
  Rev. B}\ }\textbf {\bibinfo {volume} {81}},\ \bibinfo {pages} {174429}
  (\bibinfo {year} {2010})}\BibitemShut {NoStop}%
\bibitem [{\citenamefont {Kuklov}\ \emph {et~al.}(2004)\citenamefont {Kuklov},
  \citenamefont {Prokof'ev},\ and\ \citenamefont {Svistunov}}]{Kuklov2004}%
  \BibitemOpen
  \bibfield  {author} {\bibinfo {author} {\bibfnamefont {A.}~\bibnamefont
  {Kuklov}}, \bibinfo {author} {\bibfnamefont {N.}~\bibnamefont {Prokof'ev}}, \
  and\ \bibinfo {author} {\bibfnamefont {B.}~\bibnamefont {Svistunov}},\ }\href
  {\doibase 10.1103/PhysRevLett.93.230402} {\bibfield  {journal} {\bibinfo
  {journal} {Phys. Rev. Lett.}\ }\textbf {\bibinfo {volume} {93}},\ \bibinfo
  {pages} {230402} (\bibinfo {year} {2004})}\BibitemShut {NoStop}%
\bibitem [{\citenamefont {Sirker}\ \emph {et~al.}(2006)\citenamefont {Sirker},
  \citenamefont {Weihong}, \citenamefont {Sushkov},\ and\ \citenamefont
  {Oitmaa}}]{Sirker2006}%
  \BibitemOpen
  \bibfield  {author} {\bibinfo {author} {\bibfnamefont {J.}~\bibnamefont
  {Sirker}}, \bibinfo {author} {\bibfnamefont {Z.}~\bibnamefont {Weihong}},
  \bibinfo {author} {\bibfnamefont {O.~P.}\ \bibnamefont {Sushkov}}, \ and\
  \bibinfo {author} {\bibfnamefont {J.}~\bibnamefont {Oitmaa}},\ }\href
  {\doibase 10.1103/PhysRevB.73.184420} {\bibfield  {journal} {\bibinfo
  {journal} {Phys. Rev. B}\ }\textbf {\bibinfo {volume} {73}},\ \bibinfo
  {pages} {184420} (\bibinfo {year} {2006})}\BibitemShut {NoStop}%
\bibitem [{\citenamefont {Kr\"{u}ger}\ and\ \citenamefont
  {Scheidl}(2006)}]{Kruger2006}%
  \BibitemOpen
  \bibfield  {author} {\bibinfo {author} {\bibfnamefont {F.}~\bibnamefont
  {Kr\"{u}ger}}\ and\ \bibinfo {author} {\bibfnamefont {S.}~\bibnamefont
  {Scheidl}},\ }\href {\doibase 10.1209/epl/i2006-10039-3} {\bibfield
  {journal} {\bibinfo  {journal} {Europhysics Letters ({EPL})}\ }\textbf
  {\bibinfo {volume} {74}},\ \bibinfo {pages} {896} (\bibinfo {year}
  {2006})}\BibitemShut {NoStop}%
\bibitem [{\citenamefont {Senthil}(2004)}]{Senthil2004}%
  \BibitemOpen
  \bibfield  {author} {\bibinfo {author} {\bibfnamefont {T.}~\bibnamefont
  {Senthil}},\ }\href {\doibase 10.1126/science.1091806} {\bibfield  {journal}
  {\bibinfo  {journal} {Science}\ }\textbf {\bibinfo {volume} {303}},\ \bibinfo
  {pages} {1490} (\bibinfo {year} {2004})}\BibitemShut {NoStop}%
\bibitem [{\citenamefont {Senthil}\ \emph {et~al.}(2004)\citenamefont
  {Senthil}, \citenamefont {Balents}, \citenamefont {Sachdev}, \citenamefont
  {Vishwanath},\ and\ \citenamefont {Fisher}}]{Senthil2004a}%
  \BibitemOpen
  \bibfield  {author} {\bibinfo {author} {\bibfnamefont {T.}~\bibnamefont
  {Senthil}}, \bibinfo {author} {\bibfnamefont {L.}~\bibnamefont {Balents}},
  \bibinfo {author} {\bibfnamefont {S.}~\bibnamefont {Sachdev}}, \bibinfo
  {author} {\bibfnamefont {A.}~\bibnamefont {Vishwanath}}, \ and\ \bibinfo
  {author} {\bibfnamefont {M.~P.~A.}\ \bibnamefont {Fisher}},\ }\href {\doibase
  10.1103/PhysRevB.70.144407} {\bibfield  {journal} {\bibinfo  {journal} {Phys.
  Rev. B}\ }\textbf {\bibinfo {volume} {70}},\ \bibinfo {pages} {144407}
  (\bibinfo {year} {2004})}\BibitemShut {NoStop}%
\bibitem [{\citenamefont {Batista}\ and\ \citenamefont
  {Trugman}(2004)}]{Batista2004}%
  \BibitemOpen
  \bibfield  {author} {\bibinfo {author} {\bibfnamefont {C.~D.}\ \bibnamefont
  {Batista}}\ and\ \bibinfo {author} {\bibfnamefont {S.~A.}\ \bibnamefont
  {Trugman}},\ }\href {\doibase 10.1103/PhysRevLett.93.217202} {\bibfield
  {journal} {\bibinfo  {journal} {Phys. Rev. Lett.}\ }\textbf {\bibinfo
  {volume} {93}},\ \bibinfo {pages} {217202} (\bibinfo {year}
  {2004})}\BibitemShut {NoStop}%
\bibitem [{\citenamefont {Sandvik}(2007)}]{Sandvik2007}%
  \BibitemOpen
  \bibfield  {author} {\bibinfo {author} {\bibfnamefont {A.~W.}\ \bibnamefont
  {Sandvik}},\ }\href {\doibase 10.1103/PhysRevLett.98.227202} {\bibfield
  {journal} {\bibinfo  {journal} {Phys. Rev. Lett.}\ }\textbf {\bibinfo
  {volume} {98}},\ \bibinfo {pages} {227202} (\bibinfo {year}
  {2007})}\BibitemShut {NoStop}%
\bibitem [{\citenamefont {Gell\'e}\ \emph {et~al.}(2008)\citenamefont
  {Gell\'e}, \citenamefont {L\"auchli}, \citenamefont {Kumar},\ and\
  \citenamefont {Mila}}]{Gelle2008}%
  \BibitemOpen
  \bibfield  {author} {\bibinfo {author} {\bibfnamefont {A.}~\bibnamefont
  {Gell\'e}}, \bibinfo {author} {\bibfnamefont {A.~M.}\ \bibnamefont
  {L\"auchli}}, \bibinfo {author} {\bibfnamefont {B.}~\bibnamefont {Kumar}}, \
  and\ \bibinfo {author} {\bibfnamefont {F.}~\bibnamefont {Mila}},\ }\href
  {\doibase 10.1103/PhysRevB.77.014419} {\bibfield  {journal} {\bibinfo
  {journal} {Phys. Rev. B}\ }\textbf {\bibinfo {volume} {77}},\ \bibinfo
  {pages} {014419} (\bibinfo {year} {2008})}\BibitemShut {NoStop}%
\bibitem [{\citenamefont {Zhitomirsky}\ and\ \citenamefont
  {Ueda}(1996)}]{Zhitomirsky1996}%
  \BibitemOpen
  \bibfield  {author} {\bibinfo {author} {\bibfnamefont {M.~E.}\ \bibnamefont
  {Zhitomirsky}}\ and\ \bibinfo {author} {\bibfnamefont {K.}~\bibnamefont
  {Ueda}},\ }\href {\doibase 10.1103/PhysRevB.54.9007} {\bibfield  {journal}
  {\bibinfo  {journal} {Phys. Rev. B}\ }\textbf {\bibinfo {volume} {54}},\
  \bibinfo {pages} {9007} (\bibinfo {year} {1996})}\BibitemShut {NoStop}%
\bibitem [{\citenamefont {Kumar}(2010)}]{Kumar2010}%
  \BibitemOpen
  \bibfield  {author} {\bibinfo {author} {\bibfnamefont {B.}~\bibnamefont
  {Kumar}},\ }\href {\doibase 10.1103/PhysRevB.82.054404} {\bibfield  {journal}
  {\bibinfo  {journal} {Phys. Rev. B}\ }\textbf {\bibinfo {volume} {82}},\
  \bibinfo {pages} {054404} (\bibinfo {year} {2010})}\BibitemShut {NoStop}%
\bibitem [{\citenamefont {Doretto}(2014)}]{Doretto2014}%
  \BibitemOpen
  \bibfield  {author} {\bibinfo {author} {\bibfnamefont {R.~L.}\ \bibnamefont
  {Doretto}},\ }\href {\doibase 10.1103/PhysRevB.89.104415} {\bibfield
  {journal} {\bibinfo  {journal} {Phys. Rev. B}\ }\textbf {\bibinfo {volume}
  {89}},\ \bibinfo {pages} {104415} (\bibinfo {year} {2014})}\BibitemShut
  {NoStop}%
\bibitem [{\citenamefont {Blaizot}\ and\ \citenamefont
  {Ripka}(1986)}]{Blaizot1986}%
  \BibitemOpen
  \bibfield  {author} {\bibinfo {author} {\bibfnamefont {J.~P.}\ \bibnamefont
  {Blaizot}}\ and\ \bibinfo {author} {\bibfnamefont {G.}~\bibnamefont
  {Ripka}},\ }\href@noop {} {\emph {\bibinfo {title} {Quantum Theory of Finite
  Systems}}}\ (\bibinfo  {publisher} {MIT, Cambridge, Mass},\ \bibinfo {year}
  {1986})\BibitemShut {NoStop}%
\bibitem [{\citenamefont {Capriotti}\ and\ \citenamefont
  {Sorella}(2000)}]{Capriotti2000}%
  \BibitemOpen
  \bibfield  {author} {\bibinfo {author} {\bibfnamefont {L.}~\bibnamefont
  {Capriotti}}\ and\ \bibinfo {author} {\bibfnamefont {S.}~\bibnamefont
  {Sorella}},\ }\href {\doibase 10.1103/PhysRevLett.84.3173} {\bibfield
  {journal} {\bibinfo  {journal} {Phys. Rev. Lett.}\ }\textbf {\bibinfo
  {volume} {84}},\ \bibinfo {pages} {3173} (\bibinfo {year}
  {2000})}\BibitemShut {NoStop}%
\bibitem [{\citenamefont {Mambrini}\ \emph {et~al.}(2006)\citenamefont
  {Mambrini}, \citenamefont {L\"auchli}, \citenamefont {Poilblanc},\ and\
  \citenamefont {Mila}}]{Mambrini2006}%
  \BibitemOpen
  \bibfield  {author} {\bibinfo {author} {\bibfnamefont {M.}~\bibnamefont
  {Mambrini}}, \bibinfo {author} {\bibfnamefont {A.}~\bibnamefont {L\"auchli}},
  \bibinfo {author} {\bibfnamefont {D.}~\bibnamefont {Poilblanc}}, \ and\
  \bibinfo {author} {\bibfnamefont {F.}~\bibnamefont {Mila}},\ }\href {\doibase
  10.1103/PhysRevB.74.144422} {\bibfield  {journal} {\bibinfo  {journal} {Phys.
  Rev. B}\ }\textbf {\bibinfo {volume} {74}},\ \bibinfo {pages} {144422}
  (\bibinfo {year} {2006})}\BibitemShut {NoStop}%
\end{thebibliography}%

\end{document}